\documentclass[apjl]{emulateapj}

\usepackage{xcolor}

\usepackage{bm}
\usepackage{amsmath}
\usepackage{empheq}
\usepackage{enumitem}

\received{\today}
\revised{XXX}
\accepted{XXX}

\def\te{t_{\rm E}}
\def\re{R_{\rm E}}	
\def\msol{{\rm M}_\odot}
\def\mjup{{\rm M}_{\rm Jup}}
\def\kms{\,{\rm km}\,{\rm s}^{-1}}
\def\minf{m_{\rm inf}}
\def\msup{m_{\rm sup}}

\shorttitle{Probing the Milky Way stellar and brown dwarf initial mass function}
\shortauthors{Chabrier \& Lenoble}
\graphicspath{{./}{figures/}}

\begin{document}

\title{Probing the Milky Way stellar and brown dwarf initial mass function with modern microlensing observations}

\author{Gilles Chabrier}
\affiliation{Ecole normale sup\'erieure de Lyon, CRAL, Universit\'e de  Lyon, UMR CNRS 5574, F-69364 Lyon Cedex 07, France}
\affiliation{School of Physics, University of Exeter, Exeter, EX4 4QL, UK}
\author{Romain Lenoble}
\affiliation{Ecole normale sup\'erieure de Lyon, CRAL, Universit\'e de Lyon, UMR CNRS 5574, F-69364 Lyon Cedex 07, France}

\email{chabrier@ens-lyon.fr, romain.lenoble@ens-lyon.fr }

\begin{abstract}
We use recent microlensing observations toward the central bulge of the Galaxy to probe the overall stellar plus brown dwarf initial mass function (IMF) in these regions well within the brown dwarf domain. We find that the IMF
is consistent with the same Chabrier (2005) IMF characteristic of the Galactic disk. In contrast, other IMFs suggested in the literature overpredict the number of short-time events, thus of very-low mass stars and brown dwarfs, compared with observations.
This, again, supports the suggestion that brown dwarfs and stars form {\it predominantly} via  the same mechanism. We show that claims for different IMFs in the stellar and substellar domains rather arise from an incorrect
parameterization of the IMF. Furthermore, we show that the IMF in the {\it central} regions of the bulge seems to be bottom-heavy, as illustrated by the large number of short-time events compared with the other regions. This
recalls our previous analysis of the IMF in massive early type galaxies and suggests the same kind of two-phase formation scenario, with the central bulge initially formed  under more violent, burst-like conditions than the rest
of the Galaxy.

\end{abstract}

\keywords{gravitational lensing: micro; Galaxy: bulge; stars: formation; stars: low mass; stars: brown dwarfs} 

\section{Introduction} \label{sec:intro}

The quest for an accurate determination of the stellar  initial mass function (IMF) over the entire star plus brown dwarf (BD) domain remains one of the most fundamental questions of astrophysics. Indeed, the IMF, i.e., the number of stars formed per (logarithmic) mass interval determines the energetic and chemical evolution of the universe as
well as its baryonic content. It is now widely agreed that the IMF turns over below about $\sim 0.6\,\msol$ compared with the historical Salpeter (1955) IMF (Kroupa 2001, Chabrier 2003, 2005). Similarly, the IMF seems to be reasonably similar in different environments, field,
star forming regions, young clusters as long as the conditions (mean temperature and density, large scale velocity dispersion) resemble those of the Milky Way (see e.g., Bastian et al. 2010). Only under extreme conditions of density and turbulence, as encountered for instance 
in massive early type galaxies (ETGs) or starburst regions, does the IMF seem to depart
from this universal behavior and to become more bottom-heavy (e.g., Treu et al. 2010, Conroy \& van Dokkum 2012, van Dokkum \& Conroy 2012, Cappelari et al. 2012, Barbosa et al. 2020, Smith 2020, Gu et al. 2022 and references therein). Even though no explanation can be considered as definitive yet to explain this behavior, the combined impact of unusual density and accretion-induced compressive turbulence at least provides a plausible explanation (Hopkins 2013, Chabrier et al. 2014a).

Microlensing experiments provide a powerful tool to probe the IMF, notably in the brown dwarf regime. Indeed, microlensing experiments are
independent of the usual photometric or integrated spectroscopic approaches, and of model-dependent mass-effective temperature or mass-luminosity relationships. Furthermore, one of the advantages of microlensing over photometric surveys is that only binaries with separations of less than a few AUs are unresolved, making the impact on the mass of individual objects more limited. Finally, the timescale $\te$ of a microlensing event is proportional to the square root of
the mass of the lens, $\sqrt{M}$, which favors the detection of low-mass objects, although at the expense of a cross section that also scales as $\sqrt{M}$. 

Recently, the Optical Gravitational Lensing Experiment (OGLE-III (Wyrzykowski et al. 2015) and OGLE-IV (Udalski et al. 2015, Mr\'oz et al. 2017, 2019)) has been regularly monitoring thousands of square degrees of the most stellar dense regions of the sky containing over a billion of objects. The OGLE-IV project consists of a series of long-term sky surveys covering the Galactic center, Magellanic System (Magellanic Cloud and Magellanic Bridge) and Galactic disk with over 2000 gravitational microlensing events per year, offering a unique statistical source.
We will use these new data to probe the IMF, notably its extension into the BD regime, in the Galactic disk and bulge.

\section{The Galactic bulge} \label{sec:bulge}

{The Galactic bulge (generally defined as a barred central structure at $\| l\|<10^o$, $\| b\|<7^o$), offers a unique opportunity to probe the stellar and brown dwarf {\it initial} mass function. 
It is observationally established that bulge stars are  $\alpha$-enhanced with respect to the Sun, suggesting that most of the early star formation in the inner part of the Galaxy, bulge and inner disk, occurred rapidly 
(McWilliam \& Rich 1994, Calamida et al. 2015, Clarkson et al. 2008). Indeed, the chemodynamical patterns of the bulge suggest that most of its stars formed early, in a rapid star-formation event, probably in a disk that later buckled into a boxy bar
(see Barbuy et al. 2018 for a recent review).
While the  stellar  population of the bulge indeed appears to be predominantly old ($\sim$9-10 Gyr) and approximately solar in metal abundance (Clarkson et al. 2008, Renzini et al., 2018; Hasselquist et al. 2020), however, the existence of a younger (age $\sim$ 2-5 Gyr) metal-rich ($[Fe/H]>0.2$) population, has been revealed recently (Bensby et al. 2017 and references therein, Zoccali 2019, Hasselquist et al. 2020). These findings suggest that the bulge experienced an initial starburst, followed by more quiescent star formation at supersolar metallicities in a disk until about 2-4 Gyr ago. The bulge may thus harbor (at least) 2 populations, produced by two distinct star forming episodes, identified by their  distinct $[Fe/H]$ and $[\alpha/Fe]$ values (see e.g., Barbuy et al. 2018, \S4.4, Queiroz et al. 2021). One of them has supersolar metallicity, is arranged in a bar plus a thin component out to about $\sim$5 kpc, confined in the plane. Another, $[\alpha/Fe]$-rich, metal poor component, located at $R_{\rm Gal}\lesssim 2$-3 kpc, 
has a shape close to a spheroid, higher dispersion and little or no rotation (e.g., Queiroz et al. 2021). Although its origin is not clear, it might be the result of a violent accretion phase at the early stage of the formation of the Galaxy, which triggered vigorous star formation (Queiroz et al. 2021). This is consistent with the recent analysis of the RR Lyrae stars in the bulge spheroid, with an age $\sim$13 Gyr (Savino et al. 2020). The bulge formation history will be discussed further in \S\ref{central} and \S\ref{sec:conclu}.

The most recent attempt to infer the IMF in the Galactic bulge from microlensing events is from Wegg et al. (2017), based on the dynamical model of the bulge, bar and inner disk of the MW of Portail et al. (2017b). Wegg et al. (2017) used the microlensing events released by OGLE-III (Wyrzykowski et al. 2015), which included a sample of 3718 events. These authors found that the IMF of the inner Galaxy is consistent with the one measured locally (Kroupa 2001, Chabrier 2005). These results, however, need to be examined further. Indeed, although throughout most of the stellar regime, the Chabrier (2003, hereafter C03), Chabrier (2005, hereafter C05) and Kroupa (2001, hereafter K01) IMFs are  similar,  the C05 one differs significantly from the other two ones near and below the bottom of the main sequence, i.e.,  in the BD regime. Both the C03 and K01 IMFs
 predict a much larger number of very low mass stars (VLMS) and BDs than the C05 IMF, notably near the H-burning limit (see Figure 3 of Chabrier 2005).
 We will come back to this point in \S\ref{sec-IMF}. 
 
 Determining which (if any) of the C03, K01 or C05 IMFs is correct is of prime importance for two reasons. First, this has an immediate consequence on the total census of BDs in the MW (see Chabrier 2005). Second, the inconsistency
between the observed number of BDs and the one predicted by the Kroupa (2001) IMF is often
 used as an argument to invoke a different formation mechanism between stars and BDs. In contrast, a Chabrier (2005) IMF extending smoothly from the stellar to the BD domain adequately reproduces  the observed BD distributions and BD/star ratios  of various young clusters (e.g., Damian et al. 2021) whereas the K01 IMF has a BD fraction more than twice this value (e.g., Chabrier 2005, Andersen et al. 2008, Parravano et al. 2011); this suggests a common dominant formation mechanism between stars and BDs (see Chabrier et al. 2014b for a review). The OGLE-IV microlensing observations (Mr\'oz et al. 2017, 2019) offer a unique possibility to
resolve this question, for 2 reasons. First, OGLE-IV observed many more fields and thus obtained much larger statistics. 
Indeed, OGLE-IV covers 121 fields for a total of $N_s=400\times10^6$ sources. It detected about 20,000 microlensing events in total, of which  $N_{ev}=8002$ events were retained in their final event rate  and optical depth maps.
Second, the OGLE-IV bulge observations were overall conducted at a higher cadence than OGLE-III, including about 12 deg$^2$ with cadences $\Gamma \gtrsim 1\,{\rm hr}^{-1}$, which were capable of detecting objects throughout the BD regime and, indeed, below it (Mr\'oz et al. 2017).

 \section{The mass function}
\label{sec-IMF}

The mass function was originally defined by Salpeter (1955) as the number density, $n=N/V$, of stars  per logarithmic mass interval,
$\xi(\log m)=dn/d\log m$. The mass {\it spectrum}, defined as 
the number density of stars  per  mass interval, $\xi(m)=dn/d m$, is also often, abusively, called mass function  in the literature, with the obvious relation $\xi(m)=\xi(\log m)/(m{\rm Ln}\,10)$.

In the present calculations, we will compare the microlensing results obtained with 4 different commonly used  IMFs, namely those of Kroupa (2001), Awiphan et al. (2016, hereafter A16), which is usually used in the Besan\c con Galactic synthetic model,  Chabrier (2003) and Chabrier (2005).
These IMFs are portrayed in Figure \ref{fig-IMFs} in App. B. Note that the A16 IMF is very similar to the K01 IMF below about 1 $\msol$.
The C03 or K01 IMFs start to differ significantly  from the C05 IMF only below $\lesssim 0.4\,\msol$.
The main reason for this difference, aside from the different functional forms, is that both the C03 and K01 IMFs have been calculated from the V-band 5 pc luminosity function (LF) of Henry \& McCarthy (1990). The C05 IMF  is based on a more recent observed sample in the J-band, a much more appropriate band for cool objects, of Reid et al. (2002), determined from the 2MASS 20 pc infrared luminosity function of late-type stars.  
This difference notably affects  the  normalization of  the IMF at the H-burning limit (i.e., the  star-BD boundary). The K01 IMF, for instance, predicts about 2.5 times more objects at the star-BD limit than observed in the 2MASS sample. Whereas
the difference between these 3 IMFs  only moderatly affects the stellar domain ($\gtrsim 0.1\msol$), it yields different distributions in the brown dwarf regime (see Figure 3 of Chabrier 2005).   Note that, given the small binary fraction in the BD domain ($<20\%$), the C05 IMFs for individual objects or for unresolved systems yield similar results (Figure 5 of Chabrier 2005).

In the present context of microlensing calculations, the mass spectrum $\xi(m)$ directly enters the effective probability $P_{{eff}}(m) \propto \frac{\sqrt{m}}{{\langle m \rangle}} \xi (m)$ (see Equation (\ref{Pm})). The mass function itself can be considered as a probability density function $P(m)=\xi(m)/n_{tot}$ for a lens to have a mass $m\in [m,m+dm]$, and thus a probability density $\int_{\minf}^{\msup}P(m)dm=1$, i.e., $\int_{\minf}^{\msup} \xi(m)dm=n_{tot}$, and  $\frac{1}{n_{tot}}\int_{\minf}^{\msup} m \xi(m)  dm=\langle m\rangle$, where the normalization $n_{tot}$ is determined by the total number density of starlike objects between $\minf$ and $\msup$ at a given location in the Galaxy.
 In the present calculations, we take $n_{tot}=1$, while the normalization is given by the Galactic mass density at a given point (see \S A2 in Appendix A).
 
\section{Fiducial Galactic model}
\label{model}

\subsection{Mathematical framework}
\label{math}
Our microlensing calculations proceed as in M\'era et al. (1998, MCS98), although with some differences, and are summarized in Appendix A. 
The integral (\ref{Gamma_dvar}) for the event rate is calculated with a Monte-Carlo integration method (see MCS98 (App. A48)). Each simulation was carried out with $10^7$ realizations for each field. The limits of the integral  of the optical depth and the event rate for the distance of the source (eqns. \ref{tau_dvar} and \ref{Gamma_dvar}) were chosen as ($D_{min},D_{max})=(0.8,20)$ kpc. Extending these limits is inconsequential. Indeed, the density for $D< 800$ pc is very small compared with the one of the bulge (Kiraga \& Paczy\'nski 1994, Peale 1998) and, given the exponentially decreasing disk and bulge densities, the results become essentially insensitive to   $D_{max}$ beyond 20 kpc (less than $0.1\%$ variation on the optical depth $\tau$). The transverse velocity of the lens, $v_\perp$, drawn randomly from the Monte Carlo algorithm, is
determined by the velocity distribution of the region the lens belongs to, namely the thin disk,  thick disk or  bulge. The limit value for the transverse velocity (Equation(\ref{Pv})) is taken to be $10^3\kms$. 
The minimum and maximum masses of the mass function $\xi(m)$ are chosen to be $M_{\rm min}= 0.01\, \msol$ and $M_{\rm max}=100\, \msol$, respectively. We have checked that taking $M_{\rm min}=0.001\msol$ does not change the results.

As detailed in Appendix A,
we take into account in our calculations the motion of the Sun  (see Eq.(A20)  of MCS98) and of the source star  in the determination of the lens velocity, as well as the variation of the distance of the source stars in the disk and the bulge (see Appendix A5). 
 The density of the lenses and the sources is the sum of the disk+bulge densities (see Appendix C4).
 
In very crowded fields, such as those observed toward the Galactic center, the observed objects can be the blend of several stars. This blending effect  
used to be a major source of concern for the interpretation of microlensing searches. 
Modern surveys, however, are much less sensitive to source blending. The OGLE experiments exclude the very blended events by using the selection criterion $f_s>0.1$  (i.e., more than 10\% of the baseline flux comes from the source), where $f_s$ is the blending parameter
($f_s$=1 corresponds to no blending, whereas $f_s\rightarrow 0$ corresponds to very strong blending). 
The timescales reported in Mr\'oz el. al (2017, 2019) are from their 5-parameter fitting procedure of the flux $F_i$ at time $t_i$, and thus are corrected for blending (see Mr\'oz et al. (2017, 2019 \S6) and Figure 3 of Mr\'oz et al. (2020) for details).
These blending corrections have been included in the final OGLE detection efficiency calculations, so the timescale histograms presented in their paper and used in the present paper for comparisons with our theoretical determinations, are corrected for blending.
Therefore, taking into account source blending in the simulations seems to be no longer necessary when comparing with the recent OGLE observations. 
{While highly blended events, whose blending parameter is less than $f_s<0.1$, have thus been excluded in the OGLE final samples, however, it is acknowledged by OGLE that their long-timescale events remain  affected by some bias (Wyrzykowski et al. 2015, \S5.1). This is obvious from the lower panel of Fig 11 of that paper: while $\te$ is constant for all events for $f_s>0.2$, it keeps increasing below about this value for the long-time events. In contrast, as stated by these authors, there is hardly any event with $\te$ shorter than 15 days at very small $f_s$. Then, only events longer than about 20 days remain affected by some bias. Following Wegg et al. (2017), we will thus only consider  events with blending proportion $f_s>0.2$ for the comparison between the model and the OGLE-IV all-fields data to ensure that there is almost no bias in the measured timescales (\S\ref{OGLE-all}.)

\subsection{Galactic model}

We consider a standard Galactic model, which includes a thin disk, a thick disk and a bulge. For the observations toward the Galactic center (GC), the contributions from the spheroid or halo are negligible, given their very small local normalization. The IMF of our fiducial model is based on the C05 IMF.
We stress that the aim of the present paper is not to get the best possible Galactic model, as explored, for instance, in Portail et al. (2017b) with complete 3D dynamical simulations, but to determine the accuracy of the main IMF models used in Galactic modeling. For such a study, the parameterized Galactic model described below
is sufficient.

\subsubsection{Bulge}
\label{sec-bulge}

The bulge is the central part of the Galaxy and is the inner part of the bar. 
The parameters of the bar, however, remain uncertain. Although it is known to be in the Galactic plane, its angle $\phi$ with the axis Sun-GC is uncertain. Our fiducial model is the one of Dwek et al. (1995, model G2, their Table 1) whose parameters are derived from the COBE data at 2.2 $\mu$m for the bulge density:

\begin{eqnarray}
\rho(x,y,z)&=&\rho_{0_b}\exp(-{{r_s^2}\over{2}}) \\
 {\rm with:} \,\,\,\, r_s&=&\Big[\bigl( (\frac{x}{x_0})^2+(\frac{y}{y_0})^2 \bigr)^2 + (\frac{z}{z_0})^4 \Big]^{1/4},
\label{bulge}
\end{eqnarray}
where $(x,y,z)$ indicate the three main axes of the bar ($x$ is along the bar length and points toward $(l,b)=(\phi,0^o)$). The major axis is 1.58 kpc (from the observations at 2.2\,$\mu$m) and the axis ratios $x_0:y_0:z_0$ for the bar are found to be $1:0.33\pm11:0.23\pm0.08$, but the angle is ill constrained. The normalization constant $\rho_{0_b}$  is determined from fitting the observed intensity $I(l,b)$  converted from luminosity density to mass density (see Dwek et al. (1995) for details). This model has been used by Calchi Novati et al. (2008) and Iocco et al. (2011). It should be borne in mind, however, that this model does not consider the most central part of the bulge ($|b|<3^o$) because of the unknown correction for dust absorption in this region. As will be seen in \S\ref{central}, this uncertainty may be consequential when examining the OGLE central fields.
The  bar is considered to be in rigid rotation with $\Omega_{\rm bar}=39\pm 3.5\kms$ kpc$^{-1}$ (Portail et al. 2017b) with a corotation radius $R_{\rm cor}=3.5\pm 0.5$ kpc  (Navarro et al. 2017, Portail et al. 2017b). 
The stellar mass of the bulge is taken to be $M_{\rm bulge}=1.88\times10^{10}\,\msol$ (including the presence of a 'photometric' non-axisymmetric long bar (Portail et al. 2017b))  and the angle of the bar $\phi=28^0$ (Wegg et al. 2013, 2015). This bulge model is in good agreement with the one derived recently by using OGLE-IV $\delta$ Scuti stars (Deka et al. 2022). A more thorough comparison between these two models is given in App. \ref{Deka}.

Observations by Gaia (Nataf et al. 2013) show that, in contrast to older studies, the velocity dispersion in the bulge is substantially anisotropic and depends  on the position within the bar. Based on these observations, our referee (private communication) has provided us with the velocity dispersion for 7 positions at $l\in[-6^o,+6^o]$, all at $b=-2^o$. While $\sigma_{{\rm bar}_l}(l)$ always remains  close ($\lesssim 10\%$) to the value of the Baade window, $\sigma_{BW}=110\,\kms$, the dispersion $\sigma_{{\rm bar}_b}(l)$ is found to be about 20\% lower at $l=\pm 6^o$ compared to $l=0^o$, with $\sigma_{{\rm bar}_b}\simeq 80\,\kms$, yielding an axis ratio $\sigma_{{\rm bar}_b}/\sigma_{{\rm bar}_l}\approx 0.8$ at these longitudes.
In order to take this anisotropy into account for a proper analysis of the event timescale distribution, we have linearly interpolated  the values of $\sigma_{{\rm bar}(l)}$ and $\sigma_{{\rm bar}(b)}$ as functions of $l$ in the table provided by the referee. However, we found out that, at least for the fields observed by OGLE-IV (see \S\ref{OGLE-all}), the event timescale distribution obtained with this correction remains very similar to the one obtained with an isotropic velocity dispersion, $\sigma_{{\rm bar}}=110\,\kms$ (see App. \ref{velocity}). 

\subsubsection{Thin and thick disks}

The model for the (stellar) thin and thick disks is the double exponential model of Bahcall \& Soneira (1980):
\begin{eqnarray}
\rho(R,z)=\rho_{0_d}\,\exp^{-\frac{R}{R_D}-\frac{|z|}{ H}},
\label{disk}
\end{eqnarray}
with a scale length $R_D=3$ kpc and scale heights $H=250$ pc and  $H$=760 pc for the thin and thick disk, respectively,
within the $\sim20\%$ uncertainty of the values inferred from the SDSS survey  (Juri\'c et al. 2008). 
This is similar to the model used in Portail et al. (2017b). The value $\rho_{0_d}=\rho_\odot \exp(R_0/R_D)$ is the stellar mass density normalization in the solar neighborhood, with $\rho_\odot=0.05\,\msol$ pc$^{-3}$ for the thin disk and about 1/20 this value for the thick disk (M\'era et al. 1998, Juri\'c et al. 2008), $R_0=8.2$ kpc is the galactocentric position of the Sun  (Brunthaler et al. 2011),
in agreement with the results of the Gravity Collaboration (2021), and $z_\odot=26\pm3$ pc its location with respect to the plane (Majaess et al. 2009). 
The total mass of the disk, including the gas, in this model is $M_d=5\times 10^{10}\,\msol$.

As mentioned above and detailed in Appendix A,
we take into account in our Monte Carlo calculations the motion of the Sun  and of the source star  in the determination of the lens velocity, as well as the variation of the distance of the source stars in the disk and the bulge (which can be larger than $R_0$). 
The Sun velocity with respect to the disk motion is  $U_{\odot}=11.1\kms,V_{\odot}=12.24\kms, W_{\odot}=7.25\kms$ (Brunthaler et al. 2011).
  
 The density of the lenses and the sources is the sum of the disk+bulge densities (Equations \ref{eqn_nlens} and \ref{nu_s}).
 
 The rotation velocity of the Galaxy is taken from Brand \& Blitz (1993):
 \begin{eqnarray}
 V_{rot}(R)=\Theta_0\times[1.00762(\frac{R}{R_0})^{0.0394}+0.00712],
 \label{vrot}
  \end{eqnarray}
 with a rotation velocity for the local standard of rest,  $\Theta_0=239\pm7\kms$,
 from VLBI observations of maser sources by Brunthaler et al. (2011). As shown by these authors, the value $\Theta_0=200\kms$ recommended by the IAU can be ruled out with high confidence. However, in Appendix C3, we examine the effect of using a lower value, namely $\Theta_0=220\kms$ which
 has been used in some models.
  As seen in Table 1 of App. C.3, the impact is quite modest.

The velocity dispersions around this mean velocity are well described by a gaussian distribution.  We use the radial, tangential and perpendicular velocity dispersions of the thin and thick disk ellipsoid velocities determined by Pasetto et al. (2012a, 2012b): 
 \begin{eqnarray}
 \sigma_r^{thin}&=&27.4\pm1.1 \kms,\,\,\sigma_r^{thick}=56.1\pm 3.8 \kms, \nonumber\\
 \sigma_\theta^{thin}&=&20.8\pm1.2 \kms,\,\,\sigma_\theta^{thick}=46.1\pm 6.7 \kms, \nonumber\\
 \sigma_z^{thin}&=&16.3\pm2.2 \kms,\,\,\sigma_z^{thick}=35.1\pm3.4 \kms. \nonumber
 \label{disp}
 \end{eqnarray}
The radial dependence of these disk velocity dispersions for Galactic distances interior to the Sun is taken into account as 

 \begin{eqnarray}
\sigma_{r,\theta,z}(R) =\sigma(R_0)_{r,\theta,z}\times [\Sigma(R)/\Sigma(R_0)]^{1/2},
 \end{eqnarray}
 where $\Sigma(R)$ denotes the disk surface density, as  given by the model.
 
Other Galactic models have been proposed in the literature. In Appendix \ref{depend}, we examine the impact of various parameters and of different models on the event characteristics, notably the histogram distribution. 
As seen from the table and figures in this Appendix, the parameters appear to be rather well constrained and the impact of the uncertainties in the Galactic model upon the optical depth and the microlensing event distribution  can be considered as rather modest, of the order of the observational  uncertainties. As examined in \S5.2, {\it these variations are smaller than those due to the different mass functions}. 
Similarly, the normalization of the histogram,  and thus the event rate $\Gamma$, depends significantly on the number of stars along the line of sight (l.o.s) (i.e., on the shape of the bulge, the angle $\phi$ of the bar, or the disk/bulge fraction). The proper normalization can be determined
by comparing the theoretical optical depth with the measured one. As will be examined in \S\ref{obs}, the one obtained with our fiducial model is in good agreement with this latter.

\subsubsection{Remnant stellar populations}
\label{pop} 

When doing the microlensing calculations toward the bulge, one must take into account the population of stellar remnants, i.e., bulge stars now in the form of white dwarfs (WDs), neutron stars (NSs) and black holes (BHs). Given the age of the bulge, $\sim 10$ Gyr, this  essentially concerns all stars initially born  with $m\gtrsim 1\msol$. We have considered two models for this population, namely Gould (2000) and Maraston (1998). 
Whereas the predictions for WDs and NSs are similar for these two models, they differ for the BHs: while Gould (2000) assumes a dispersion around 5 $\msol$, Maraston (1998) takes masses in the range 20-50 $\msol$. It is now well determined that BHs have typical masses around $10\,\msol$
(e.g., Sahu et al. 2022). Black holes, however, represent a negligible fraction of starlike objects ($<1\%$) so their impact on the event (mass) distribution is negligible.

\subsubsection{Binaries} 

The typical Einstein radius of microlensing events toward the bulge is $\sim 2$ AU. Binaries with smaller separations are not resolved
and thus affect the mass determination of the lens and thus of the IMF. Such events generally cannot be fitted accurately by the single lens model and have been removed from the OGLE sample (Wyrzykowski et al. 2015). The fraction of events affected by this bias is about $6\%$
(Sumi et al. 2013), which yields a factor $1.09$ on the optical depth (Sumi et al. 2013, Mr\'oz et al. 2019). The impact on the $t_{\rm E}$-histogram (as well as various other specific biases) has been estimated to be $\lesssim 10\%$ (Glicenstein 2003). Performing a more
detailed analysis based on population synthesis, Wegg et al. (2017) found out that the correction due to binaries does not provide a suffcient information to distinguish the different IMF signatures.
 
\section{Comparison with observations}
\label{obs}

We have compared our calculations, performed with $10^7$ realizations for each field in every simulation, to the microlensing results obtained in the OGLE-IV observations (Mr\'oz et al. 2019). The latter cover 121 fields located
toward the Galactic bulge ($|b|\le 7^o, |l|\le 10^o$) for a total of about 160 deg$^2$,
and a total exposure of $E=N_s\times T_{obs}=(400\times10^6)\times 8=32\times10^8$ star-yr, revealing 8000 microlensing events in their final event rate  and optical depth maps (Mr\'oz et al. 2019). OGLE-IV has superseded the results obtained previously with OGLE-III
(Wyrzykowski et al. 2015), which detected 3718 events for a total exposure of 1.2 $\times10^8$ star-yr. Furthermore, OGLE-III does not provide the number of monitored stars  in the fields, precluding a determination of the rate of events and thus an accurate comparison of the observed and theoretical event distributions.
The data and the efficiencies were kindly provided by Przemek Mr\'oz\footnote{Note that 3 fields (BLG535.30-32) have been removed in  Table 5 of Mr\'oz et al. (2017) and thus should not be listed in their Table 3. Furthermore, the weight (=1/efficiency)
in the first online version of Table 3 was incorrect, giving different efficiencies between the 2017 and 2019 papers for the same data. This has now been updated (P. Mr\'oz 2021, private communication).}. We have also made comparisons with the results of the (revised) MOA-II survey (Sumi et al. 2016), which detected 474 events for a total exposure of 0.22 $\times 10^8$ star-yr. It must be noted that these observations do not represent a comprehensive list of OGLE-IV events, as they do not include the central-most Galactic fields. The latter (observed with higher cadences) have been published separately (Mr\'oz et al. 2017) and will be examined in \S \ref{central}. Similarly, they do not include the OGLE-IV events in the Galactic plane (Mr\'oz et al. 2020), to be examined in \S \ref{plane}.
Recently, the  first catalog of Gaia microlensing events from all over the sky was released (Wyrzykowski et al. 2022). They detected 363 events and the comparison of timescales reveals generally good agreement with the measurements of OGLE mentioned above. However, besides the low statistics, the sample is found to be significantly incomplete for the bulge region ($\lesssim 30\%$), notably for short timescales and thus cannot presently be used  for detailed comparisons.

In order compare our model with the truly observed data, the experimental efficiency is straightforwardly applied  to our calculations  with a rejection algorithm.

 \subsection{Optical Depth, Event Rate and Mean Characteristic Timescale}
 \label{tau-gamma}

 Figure \ref{fig-tau} displays the optical depth $\tau$, the event rate $\Gamma$ and the average characteristic timescale $\langle  \te \rangle$ obtained using our fiducial model for 3 different values of the $\beta$ parameter ($\beta=-1.0, -1.1$ and -1.2)
  for the density of sources stars (eq.(\ref{nu_s})), within the range $-3\le \beta \le -1$ suggested by Kiraga \& Paczy\'nski (1994). A lower value of $\beta$ means fainter stars, thus a lower number of detectable stars, decreasing the number of events (Equation \ref{Gamma_dvar}).  The results are compared to those determined by OGLE-IV (OGLE-III did not determine the optical depth) and MOA II as a function of latitude, $b$. The data are averaged values for $|l|<3^o$, with bins $\Delta b=0.6^o$. OGLE-IV still has a systematic error of about $10\%$ on the estimation of the number of sources, which dominates the reported error bar of the measurement. 
As discussed in Mr\'oz et al. (2019), the difference between the OGLE-IV and MOA II optical depth determinations, which reaches up to $\sim 30\%$, stems most likely from the determination of this source number. As seen in the figure, for $|b|\le 2^o$, i.e., the central fields, the strong absorption of the ISM strongly affects the detection, significantly decreasing  the efficiency. Note that OGLE-IV only considers the events shorter than $\te$=300 days, and so implies an efficiency $\epsilon(t)$=0 above this value. 

Overall, the agreement between our model and the observations is very good, except for the very central fields. We will come back to this point in \S\ref{central}. We have verified that a
value $\beta<-1.2$ significantly underestimates  $\tau$, $\Gamma$ and the $\te$-distribution.
Note that the fact that $\Gamma$ is underestimated for the central fields with a C05 IMF (see Fig. \ref{fig-tau}) is not surprising since, as examined in \S\ref{central}, the IMF in the central fields departs for this IMF.
As discussed in  Mr\'oz et al. (2019), the mean timescales of microlensing events in the Galactic bulge increase with Galactic latitude, with shorter average values closer to the Galactic plane, which is in agreement with theoretical expectations (see their \S8.1).

The number of observed events sharply decreases at low
Galactic latitudes ($|b|\le 1^o$) owing to extremely large interstellar
extinction. Source stars of events detected in
this region are located closer than those at larger Galactic
latitudes; hence, the number of potential lenses (in the optical), and thus the
optical depth is smaller (Mr\'oz et al. 2019).

We have also explored the dependence of the microlensing optical depth and event rate obtained with our fiducial model as a function of
Galactic longitude, and compared with the OGLE-IV data in the Galactic plane (Mr\'oz et al. 2020). This study will be presented in detail in \S\ref{plane}.

\begin{figure}[h!]
\vspace{1cm}
\center{
\includegraphics[height=5cm,width=6cm] {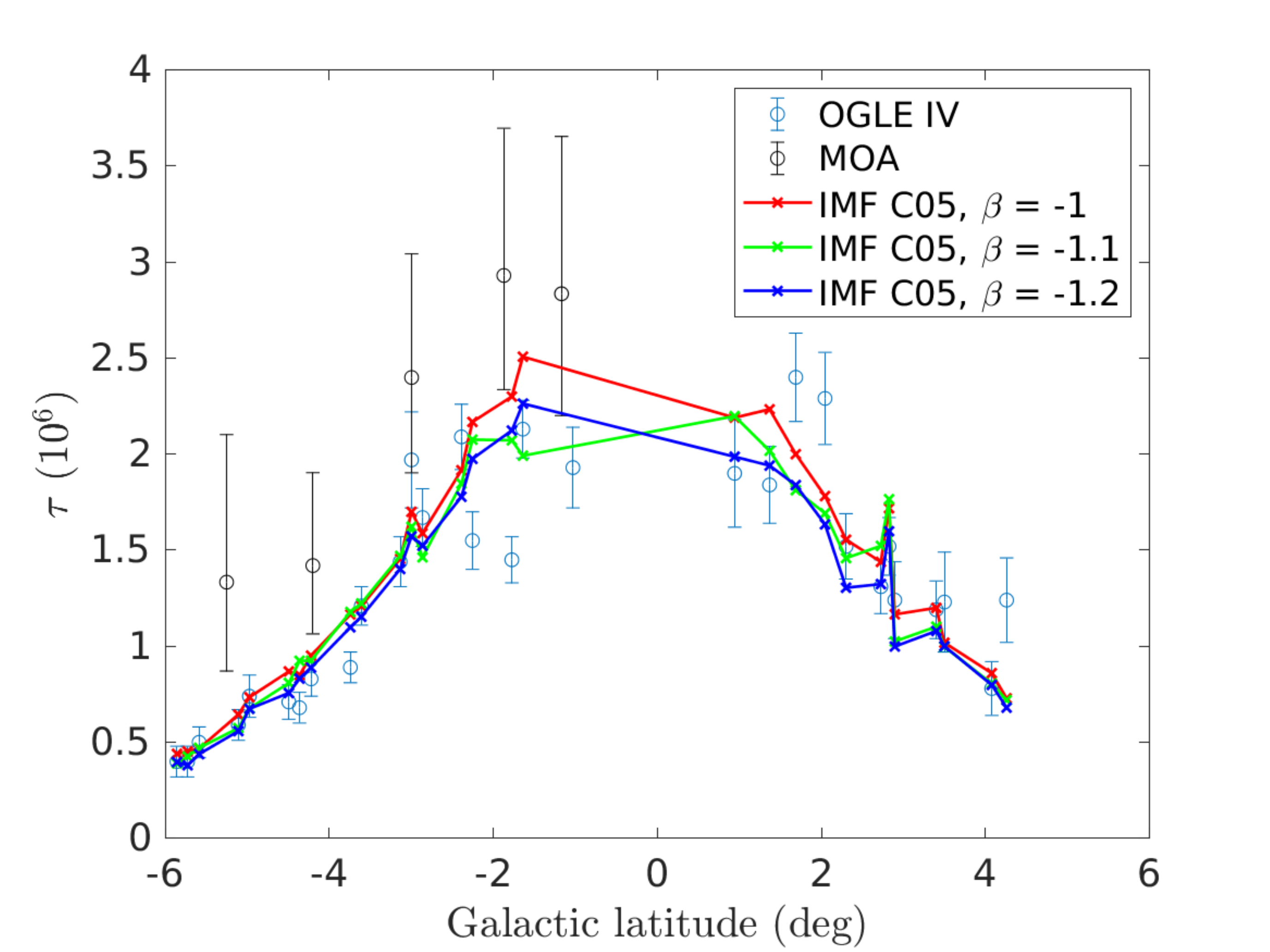} 
\includegraphics[height=5cm,width=6cm] {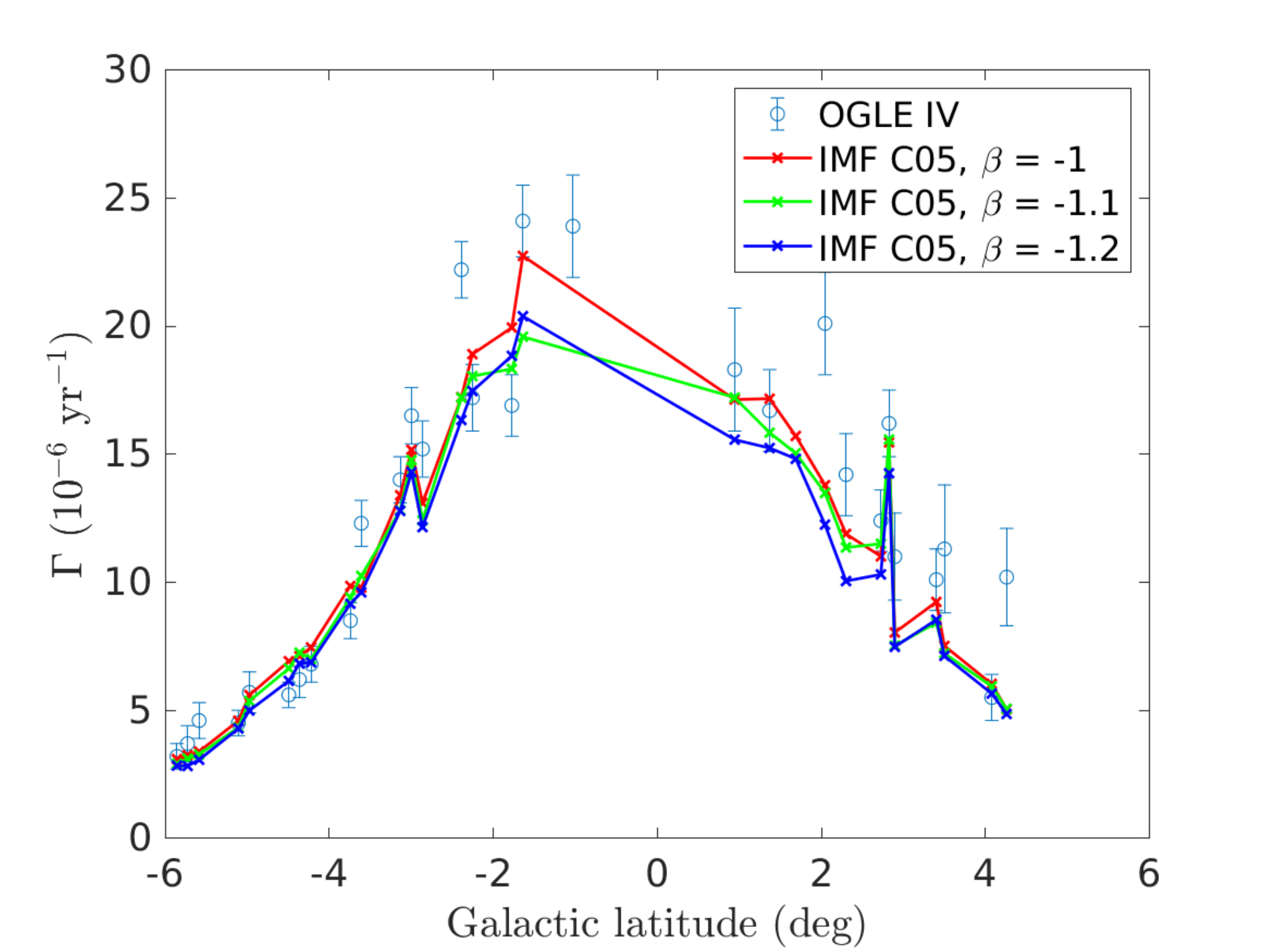} 
\includegraphics[height=5cm,width=6cm] {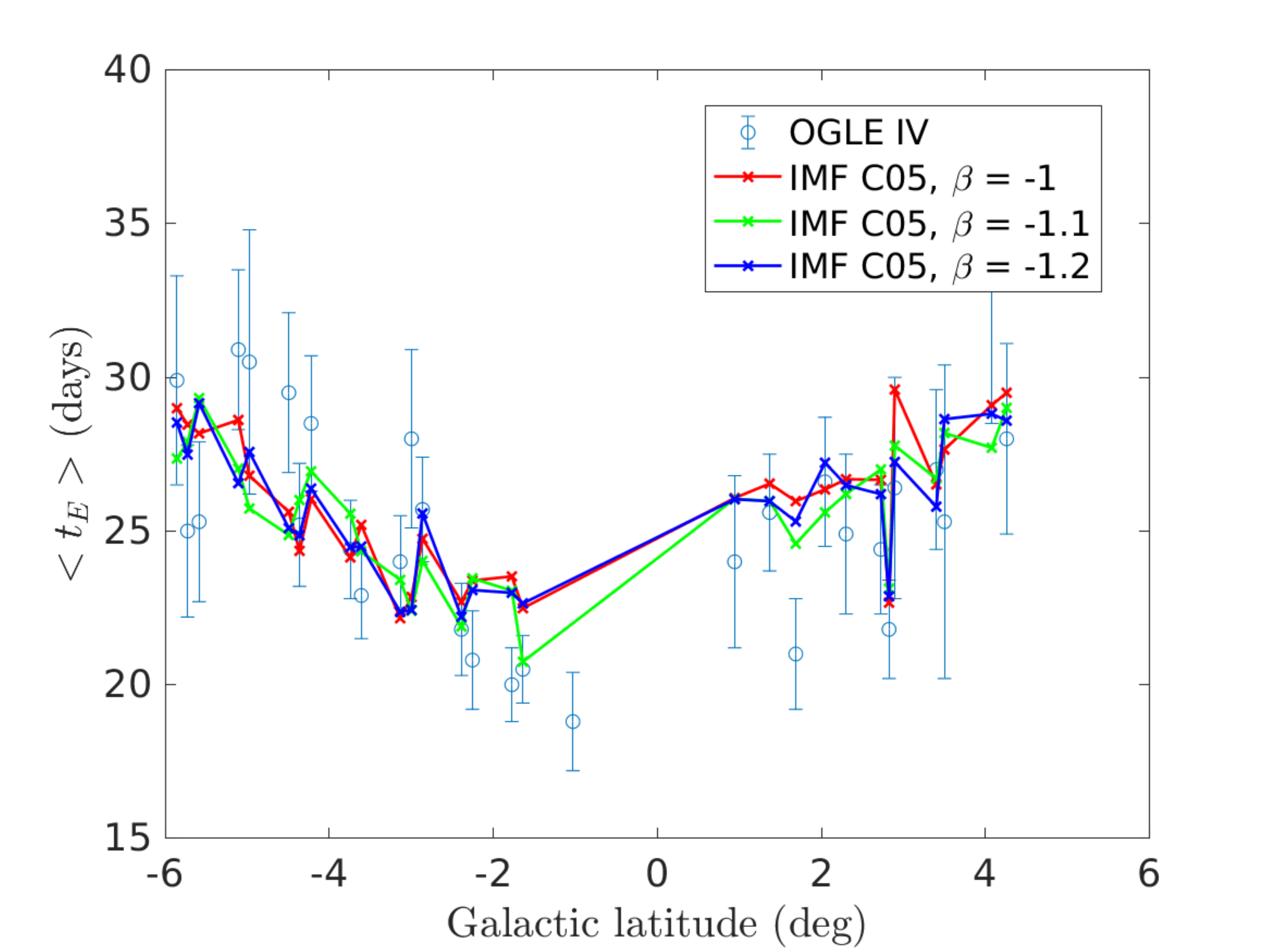} } 
  \caption{Comparison of  $\tau$, $\Gamma$ and $\langle t_{\rm E}\rangle$ obtained with our fiducial model  with the MOA II and OGLE-IV experiments for all stars as a function of latitude $b$, for 3 values of the parameter $\beta$ in eq.(\ref{nu_s}). The data are averaged for $|l|<3^o$ and binned by $\Delta b$=0.6$^o$. The efficiency of each corresponding OGLE-IV field has been applied, which explains the 'spikes' at some latitudes.}
\label{fig-tau}
\end{figure}

 \begin{figure}[h!]
\vspace{0cm}
\center
{\includegraphics[height=9cm,width=9.4cm] {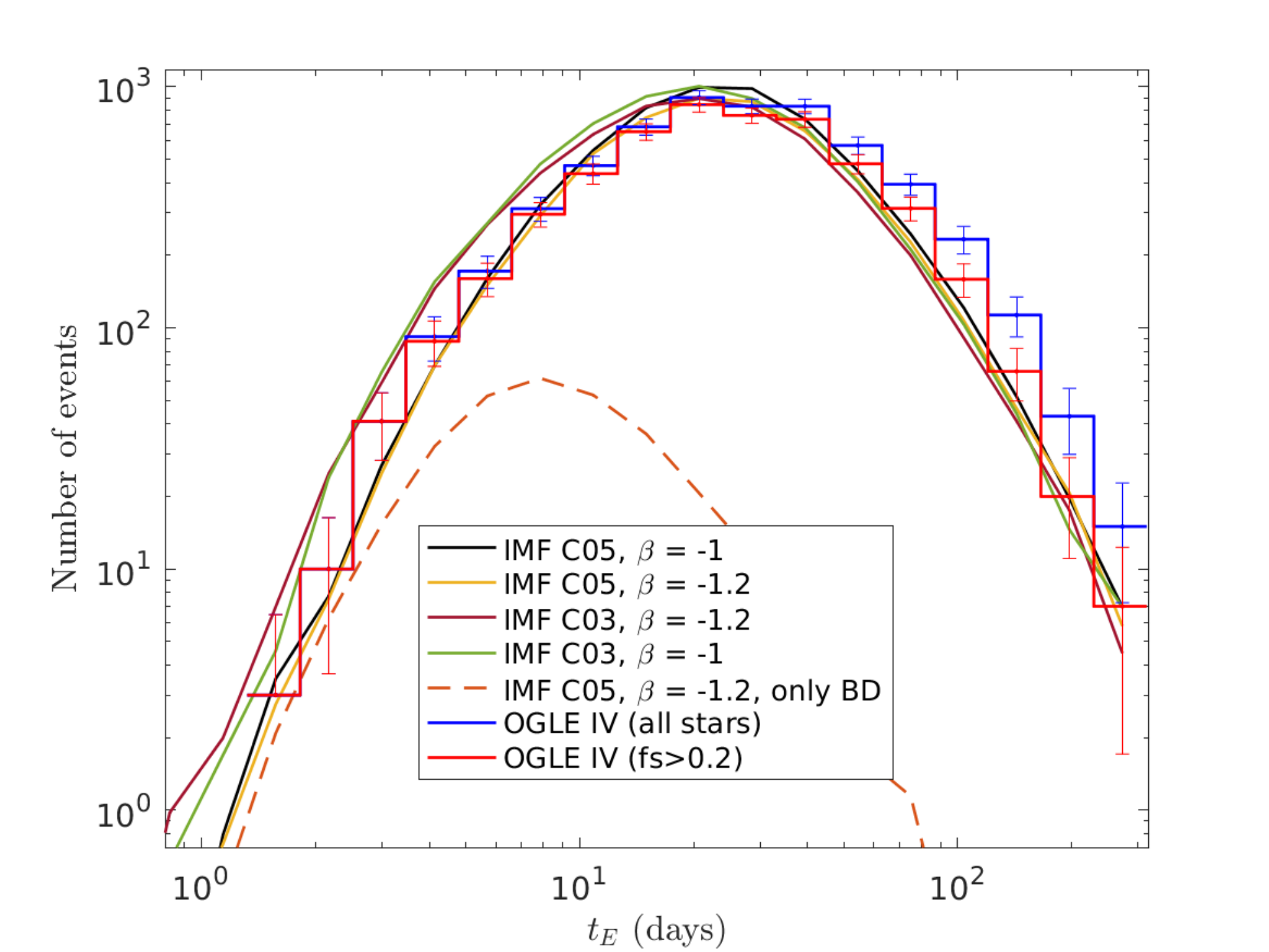} 
\hspace{-1.1cm}
\includegraphics[height=9.cm,width=9.4cm]{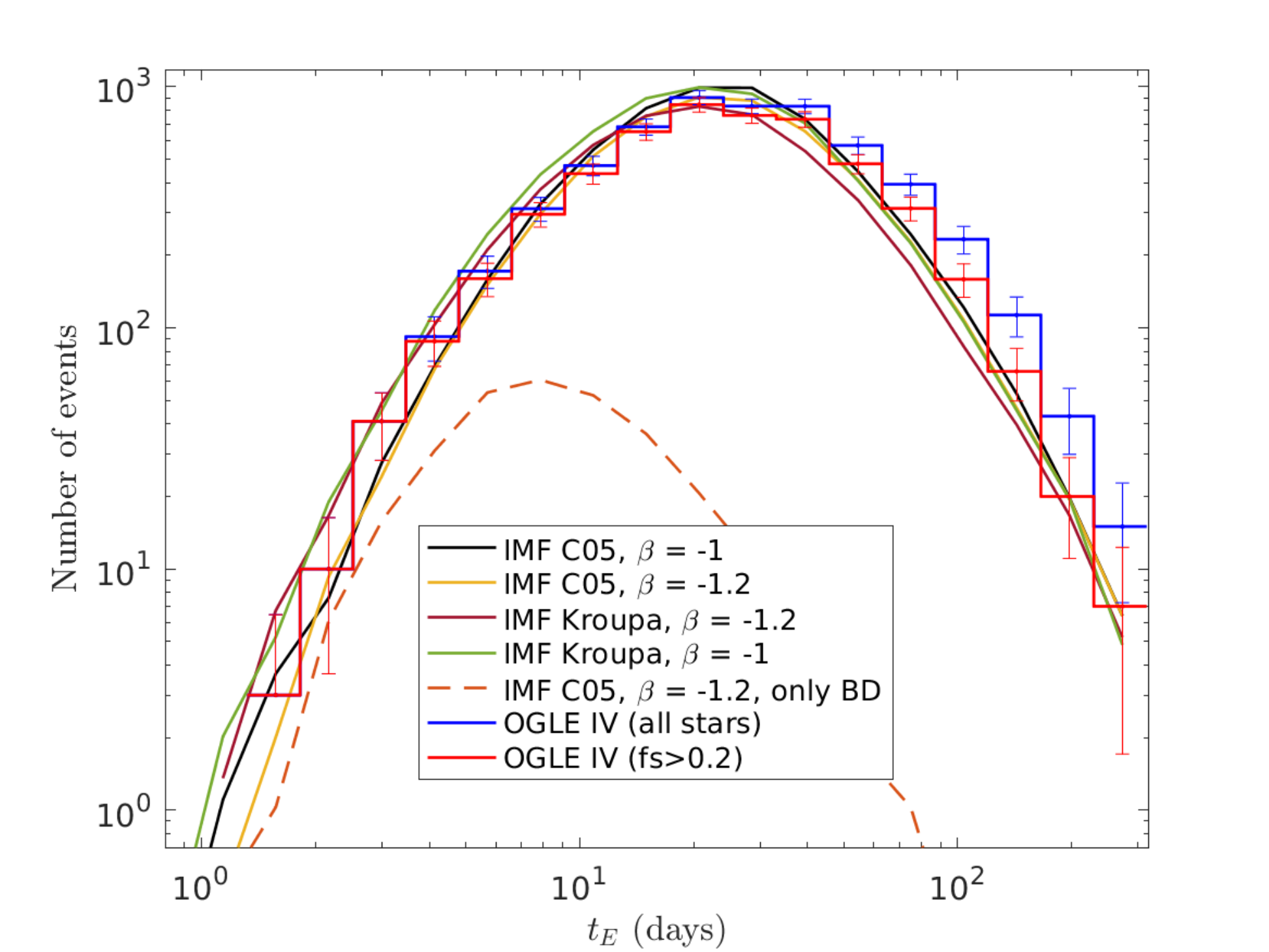} 
} 
\vspace{0cm}
  \caption{
 Comparison of  the $t_{\rm E}$-histograms obtained with our fiducial Galactic model, for 3 IMFs (C05, C03, K01), for $\beta=-1.0$ and -1.2 for the density of sources stars (eq.(\ref{nu_s})), with the results of OGLE-IV all-fields observations. The efficiency of each corresponding field has been applied. The lower dashed line shows the contribution from brown dwarfs. The bin size is the same as in Mr\'oz et al. (2019), namely, 25 bins equally spaced  between $\log \te =0.1$ and 2.5.}
\label{fig-compar}
\end{figure}

 \subsection{Time Histograms. Probing the Star+BD IMF}
 
 \subsubsection{OGLE All Fields}
 \label{OGLE-all}

 \begin{figure}[h!]
\vspace{0cm}

\center{\includegraphics[height=8.5cm,width=9.5cm]{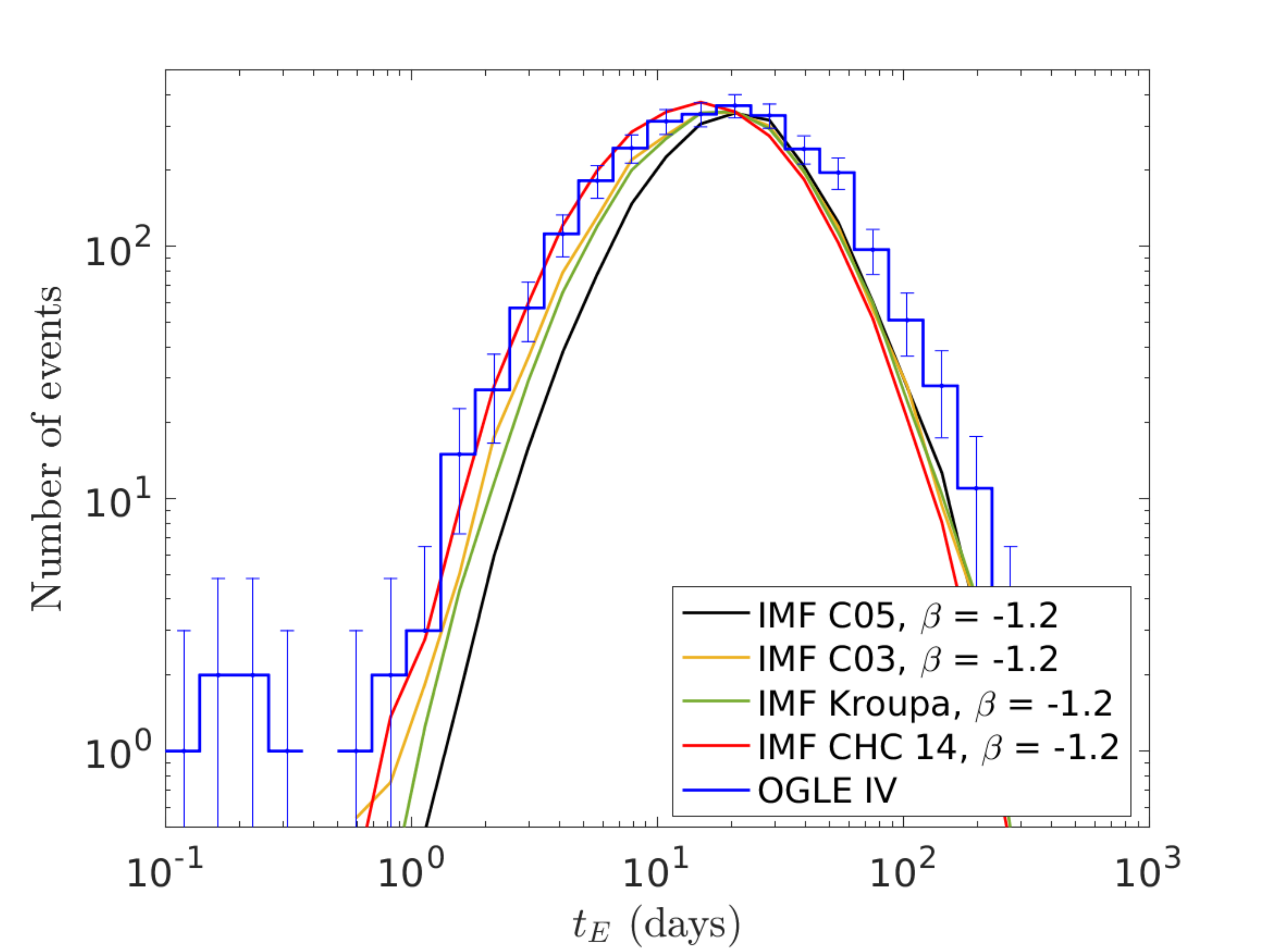} }
\vspace{0cm}
  \caption{Same as Figure \ref{fig-compar} for the OGLE-IV central fields (Mr\'oz et al. 2017) as obtained with the same IMFs as in Figure \ref{fig-compar} and the one portrayed in Figure \ref{fig-IMFs} in App. B, similar to the ones derived in Chabrier et al. (2014a, CHC14) (see text).}
\label{fig-central}
\end{figure}

 Before going further in the comparison between the model and the data, it should be noted that both the OGLE-III (Wyrzykowski et al. 2015) and OGLE-IV (Mr\'oz et al. 2019) individual events analyses are based on the basic Paczynski (1996) microlensing model. This model ignores the motion of the
Earth around the Sun. The Paczynski (1996) parameters are thus heliocentric in nature but estimated in the geocentric frame. The Earth's parallax effect, caused by variable magnification due to Earth's motion around the Sun, must then be taken into account for a proper analysis of the $\te$-histograms (Gould 2004). Most of the bulge microlensing events, however, are short (less than a couple of months) so the Earth's motion can be ignored. This is no longer true for long-time events. Such a reanalysis of the OGLE-III and OGLE-IV data was recently conducted by Golovich et al. (2022). 
As shown by these authors, the Earth's parallax correction decreases $\te$ for the long-time events, yielding a distribution similar to the one displayed in our Figure \ref{fig-compar} for $f_s>0.2$  (see their Figure 12).

 Figure \ref{fig-compar} compares the results obtained with our fiducial model (C05 IMF) and  the C03 and K01 IMFs to the data,
for 2 values of the $\beta$  in Equation(\ref{nu_s}), namely $\beta=-1.0$ and -1.2. The Awiphan et al. (2016) IMF yields results similar to K01 and thus is not displayed in the figure.
We  show the distribution as a function of $\log \te$, which highlights the short times and is less prone to statistical fluctuations of rare short-time and long-time events. As mentioned in \S\ref{math}, only events with a blending proportion $f_s>0.2$ should be considered for the comparison between the model and the data to ensure there is almost no bias in the measured timescales (orange curve in Figure \ref{fig-compar}).

{As seen in the figure, the agreement model-observations with our fiducial model (C05 IMF) can be considered as fairly good, well within the uncertainties of the global Galactic modeling (see Appendix \ref{depend}). A value $\beta=-1.2$ yields a nearly perfect agreement with the data.
In contrast, both the C03 and K01  (or similarly A16) IMFs fail to reproduce the correct histogram.
These IMFs yield a significant excess of events at and below the peak region and substantially
overestimate the number of events over the $\sim 5$-25 day domain, i.e., the very low mass star and BD domains. Conversely, they tend to
underestimate the number of large-timescale events and would require a substantial modification
of the parameters of the Galactic model in order to agree with the data. We also note that while the fiducial model properly reproduces the location of the peak, the C03 and K01 ones are shifted toward shorter timescales, a direct consequence of the shift of these IMFs towards smaller masses compared with C05 (see Figure \ref{fig-IMFs} in Appendix \ref{massfunc}).
As mentioned earlier and explored in detail in App. \ref {depend}, these differences in the timescale distributions are larger than the ones due to uncertainties in the various model parameters. Therefore, even though one must remain cautious with (statistical) microlensing analysis, it seems difficult to explain
such a disagreement with model uncertainties. Similarly, to reconcile these predicted timescales with the observations would imply that the detection efficiency and/or the blending correction are significantly either over- or underestimated, depending on the timescale range (see Equation (\ref{gammaexp})), which is at odds with the detailed analysis carried out in Mr\'oz et al. (2019) (see their \S5-7). At the location of the peak (around $\sim20$ days), we have verified that the fractions of bulge-bulge, bulge-disk and disk-disk events amount to about $\sim$ 70\%, 30\% and $<1\%$, respectively. Only for events
with $\te\gtrsim 100$ days does the number of disk-bulge events start to dominate the bulge-bulge one. The figure also displays the contribution from BD events. As seen, these latter start to contribute significantly ($\gtrsim 50\%$) below $\lesssim 6$ days (see Equation (\ref{teobs})).

 \subsubsection{The OGLE central fields}
 \label{central}

Between 2010 and 2015, Mr\'oz et al. (2017) observed the 9 central fields of OGLE-IV with the highest-cadence (BLG500, BLG501, BLG504, BLG505, BLG506, BLG511, BLG512, BLG534 and BLG611), which include 2617 detected events. These  fields have a better efficiency for the very short-time events. The global histogram of these fields is portrayed in
Figure \ref{fig-central}. Assuming that the same $f_S>0.2$ correction as for the other fields applies (we do not have these data for these fields), the long timescale tail should be well reproduced with the model (Fig. \ref{fig-compar}). In contrast, it is clear that the model
severely underestimates the number of events below $\te\lesssim$ 20 days, for all of the IMFs considered previously. 
Therefore, for these OGLE-IV {\it central fields}, the model underestimates the number of short-time events and yields larger Einstein timescales compared with observations. 
It is worth stressing that all central fields exhibit similar distributions, excluding the possibility that the peculiar global distribution of the central fields could be due to a single atypical field. 
Therefore, there is definitely an excess of short-time events, as clearly seen in the figure: the central fields yield a  value  for $\langle \te\rangle$ that is smaller not only than the model predictions but also than the other OGLE-IV fields (see Figure \ref{fig-compar}), as already noted by Mr\'oz et al. (2019). 
We verified that using a lower limit $\minf=0.001\msol$ for the IMF does not change the results. 

Such a difference between the central and peripheral fields can have several possible explanations. (1) The efficiency toward the central fields is significantly overestimated (see \S8.1 and Figure 11 of Mr\'oz et al. 2019); (2) the blending is severely underestimated.
However, as discussed in \S\ref{math}, blending is unlikely to be the source of the difference. (3) Several studies (see e.g., Lian et al. 2020a and references therein) have revealed a complex range of stellar populations in the bulge and complex structures in their chemical abundances and kinematics, pointing to several phases of star formation history for the inner Galaxy within $r_{\rm GC}<3$ kpc (Lian et al. 2020b). Assuming one single velocity dispersion might then be a too simplistic approach in this region. (4) Incompleteness in stellar number counts, which increases toward the GC, might also lead to an overestimate of $\tau$ and $\Gamma$ (see e.g., Sumi \& Penny 2016). As noted by these authors, the incompleteness increases at lower $|b|$ because of the higher stellar density and the higher interstellar extinction. Indeed, there are several pieces of evidence for a stellar overdensity in the plane near the GC (Launhardt et al. 2002, Nishiyama et al. 2013, Sch\"onrich et al. 2015, Debattista et al. 2015, see Portail et al. 2017b and references therein). 
Note, e.g., the large difference in $\tau$ and $\Gamma$ between 2
juxtaposed fields when one of the two belongs to the central fields (e.g., BLG505 and BLG513, see Table 7 of Mr\'oz et al. 2019).

Finally, if all of the aforementioned possible sources of bias are excluded as a possibility to resolve this issue, the observed timescale histogram might reveal a
genuine difference in the microlensing event distributions between the peripheral and central fields, yielding potentially shorter events in the latter. As seen from Equation (\ref{te}), this can stem from 3 different causes in the central part of the bulge, namely: (i) a higher lens-source proper motion, transverse velocity $v_\perp$, (ii) a smaller relative lens-source distance, i.e., a smaller $D_Sx(1-x)$, (iii) a genuine excess of very low mass objects, due to peculiar star formation conditions. Naively, one expects items (i) and (ii) to affect all events, not preferentially short ones, leaving the shape of the event distribution unchanged. By construction, these effects are taken into account in our Monte Carlo calculations (eqns.(\ref{Pm})-(\ref{Pv})). We have carefully verified this issue in Appendix \ref{central-fields}. As shown in this appendix, the statistical distributions of the number of events as a function of $v_\perp$ and $D_Sx(1-x)$ for the central field conditions are similar for the 4 different IMFs examined in Figure \ref{fig-central}, confirming the fact that the effective probabilities $P_{{eff}}(x)$ and $P_{{eff}}(v_\perp)$ (eqns.(\ref{Px}),(\ref{Pv})) do not depend on the mass. In contrast, the timescale distributions differ significantly between different IMFs.
The atypical event timescale distribution in Figure \ref{fig-central} might thus truly stem from a different, bottom-heavy IMF. In order to test this hypothesis, we have calculated the histogram of the central fields with an IMF corresponding to conditions somewhere between the Milky Way conditions and "Case 2" in Table 2 of Chabrier et al. (2014a) (see Figure \ref{fig-IMFs} in App. B). The result is portrayed in Figure \ref{fig-central} with the denoted CHC14 IMF (to be understood as a bottom-heavy type IMF as described in Chabrier et al. 2014, Table 1 and Fig. 1). Our Galactic model with this type of bottom-heavy IMF yields a significantly better agreement with the data, notably the short timescales. {Making comparisons with {\it each} of the 9 aforementioned central fields, we have verified (keeping in mind the low statistics for each of these fields) that the timescale distributions for all the fields located at $b=-2^0,|l|< 2^o$ are consistent  with a CHC14-type IMF, i.e. a bottom-heavy IMF compared with the C05 one. Moving outward around this region, the IMF smoothly transits from a CHC14-type to C03 (itself bottom-heavy compared with C05) to C05 in the peripheral fields.

 As mentioned in the Introduction, such a peculiar IMF has been advocated for the progenitors of massive ETGs and has been suggested to be due to the high density and (accretion-induced) turbulence, and thus high external pressure and surface density (thus compactness), during the bursty formation stage of these galaxies (Hopkins 2013, Chabrier et al. 2014a, Barbarosa et al. 2020). The formation history of the central part of the Galaxy indeed occurred within $\sim 0.5$ Gyr, i.e., under a burst-like mode, and thus differs significantly from the one of the local disk. 
 Analysis of the APOGEE and Gaia data has recently assessed the existence of an accreted structure located in the inner Galaxy that likely occurred in the early life of the MW (Horta et al. 2021), so it is not implausible that part of the bulge stellar population originates from these early events. Another argument in favor of such an early accretion event is the evidence for two separate components  in the RR Lyrae population, with distinct spatial distributions and marginally different kinematics, one population being  centrally concentrated (e.g., Savino et al. 2020). A possible interpretation is that this population was born prior to bar formation, as its spatial location, kinematics, and pulsation properties suggest possibly an accretion event in the early life of the Galaxy (Kunder et al. 2020, Du et al. 2020).
It is not inconceivable that such star formation conditions would be more similar to the ones encountered in ETGs than under quiescent conditions such as in the MW. As shown in Chabrier et al. (2014a), in a gravo-turbulent scenario of star formation, these uncommon conditions can indeed lead to bottom-heavy IMFs.

The 5 shortest-time events, $0<\te<0.5$ day (Mr\'oz et al. 2017), displayed in Figure \ref{fig-central}, are most likely events due to lenses of the order of a Jupiter mass or less either ejected or on large orbits (Mr\'oz et al. 2017, 2020),  as confirmed by Gould et al. (2022), and thus are not representative of the population described by the IMF. 
It is worth stressing that the nondetection of a large number of short-timescale events in these 2 experiments strengthens the absence of a large population
of free-floating or wide-orbit Jupiter-mass planets, in contrast to previous claims (Sumi et al. 2011).  According to this analysis, about 5\% of such objects around main sequence stars could explain these statistics (Mr\'oz et al. 2017, 2020).

\subsubsection{OGLE Galactic plane}
\label{plane}

As part of the OGLE-IV survey, the OGLE GVS one (Mr\'oz et al. 2020) was carried out during 2013-2019. The fields are located along the Galactic plane 
($|b|\le 7^o, 0^o<l<50^o, 190^o<l<360^o$)
and in an extended area around the outer Galactic bulge. They cover an area of about 2800 deg$^2$ and contain over 1.8$\times 10^8$ sources and 630 detected  events.
Figures 7 and 8  of Mr\'oz et al. (2020) present the detection efficiency-corrected distributions of event timescales
in the Galactic plane fields ($l>20^o$).
As noted in that paper, these histograms have a similar shape (slopes of short- and long-timescale tails) to those in the central Galactic bulge but events in the  disk are longer. 
 In particular, their sample contains only two events with $t_{\rm E} < 10 $ days
at $l>20^o$, with timescales of about 5.7 and 7.2 days (see \S6.2 and Figure 7 of Mr\'oz et al. (2020)).
The Einstein timescales of Galactic plane 
events are, on average, three times longer than those of Galactic bulge events, with
little dependence on the Galactic longitude. This property is expected from the
theoretical point of view because lensing objects are closer than those
toward the Galactic bulge (so their Einstein radii are larger). Moreover, as the
observer, lens, and source, {\it all located in the Galactic disk}, are moving in a similar
direction, the relative lens-source proper motions should be lower than those in the Galactic bulge.

For sake of completeness in our study, we have also compared our fiducial model calculations
with these data. Figure \ref{fig-compar-long} displays the optical depth and event rate obtained with our fiducial model as a function of longitude. For a  fixed longitude,
the results are averaged over 10 values of the latitude  ($-7^o\le b\le +7^o$, as in Mr\'oz et al. 2020) and the longitude is sampled every $10^o$ at the same interval. 
Figure  \ref{fig-compar-lat} portrays the same kind of comparison as a function of latitude for
10 averaged longitudes. The sampled latitude is averaged over 50 angles between $240^o$ and $330^o$,
as in Mr\'oz et al. (2020).

 \begin{figure}[h!]
\vspace{0cm}
\center{
\includegraphics*[height=5.8cm,width=6.cm]{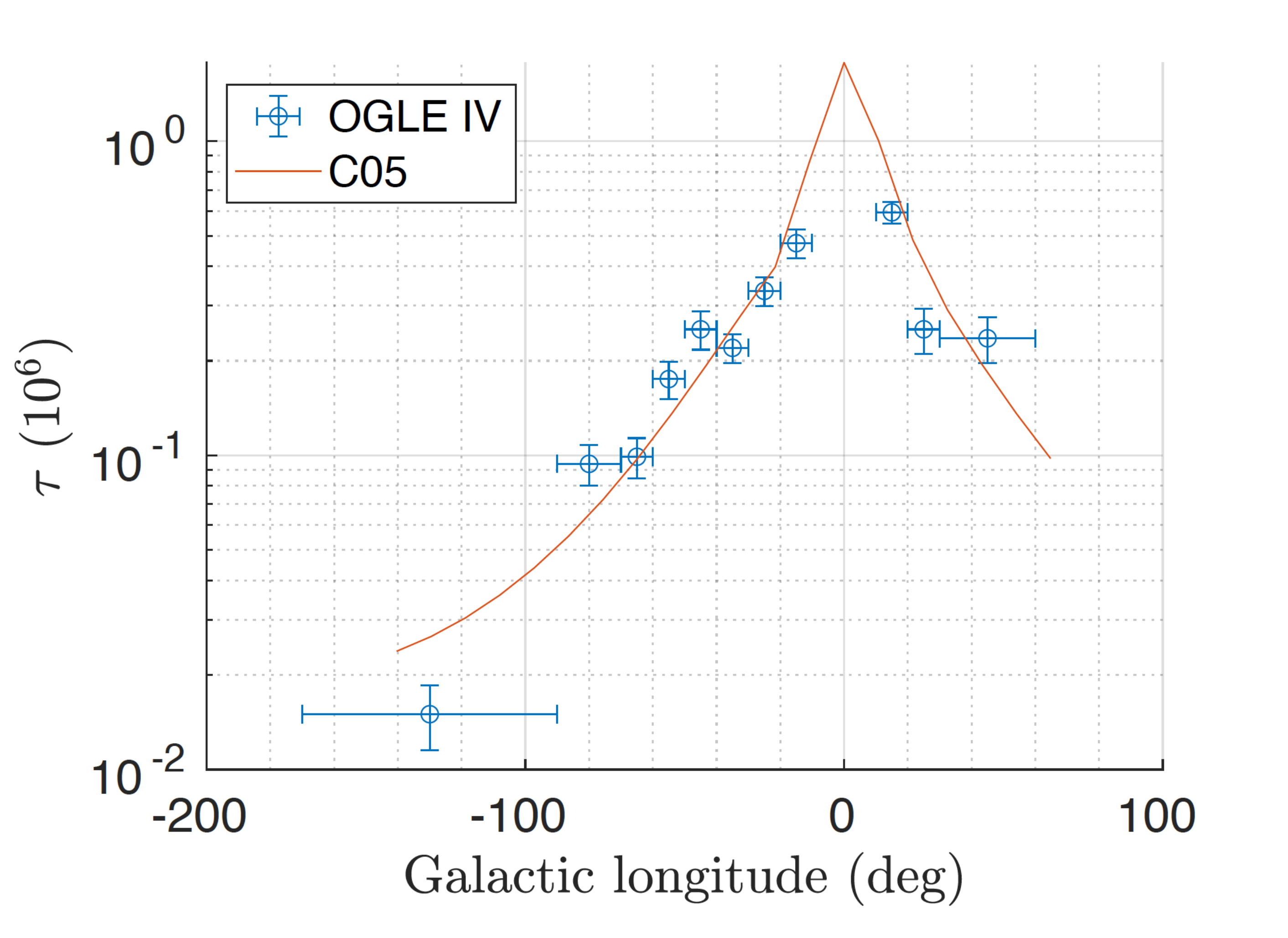}
\includegraphics*[height=5.8cm,width=6.cm]{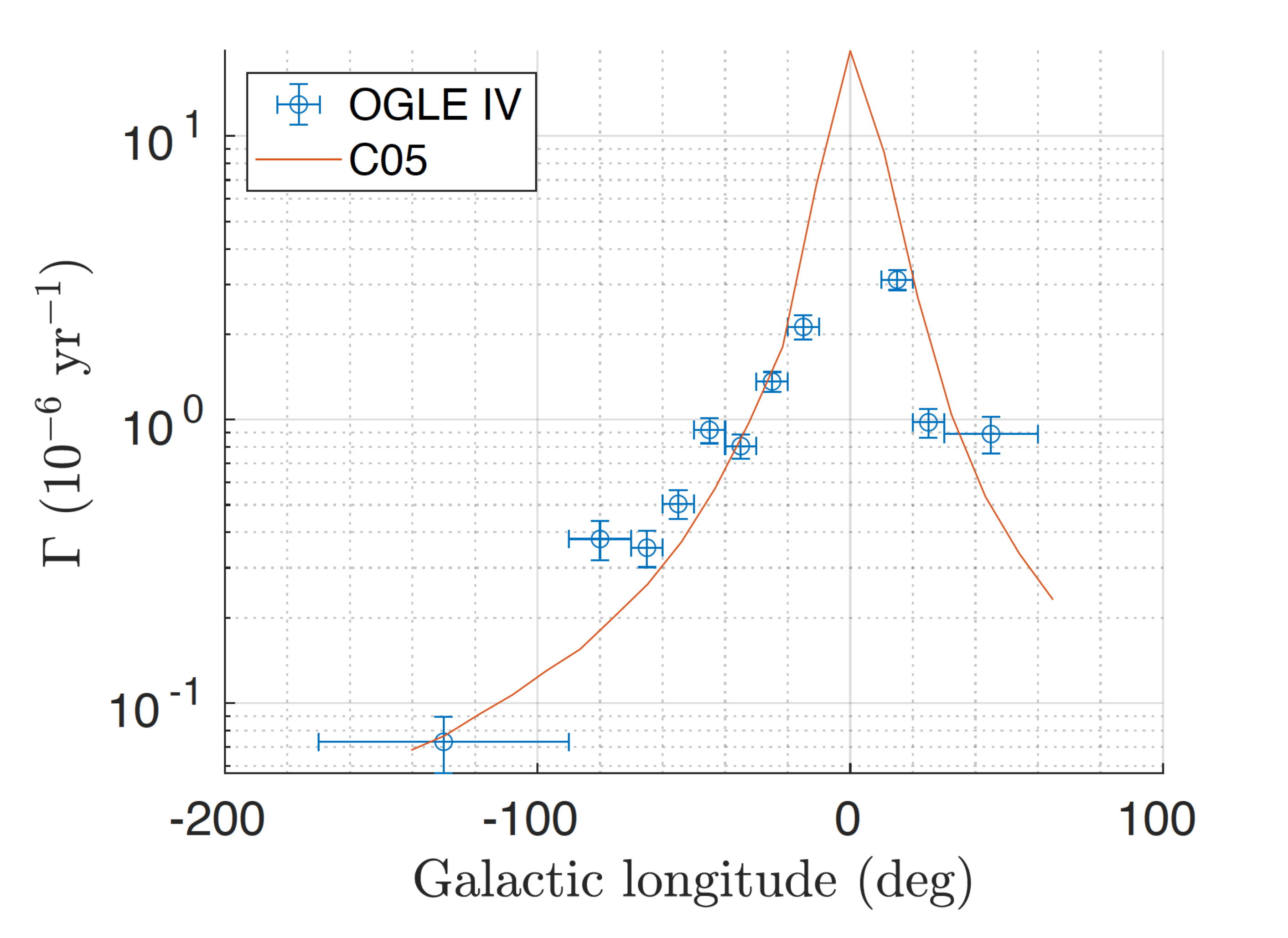}
}
  \caption{Averaged optical depth (top) 
  and event rate (bottom) 
  distributions in the Galactic plane as a function of longitude. 
  Data: Mr\'oz et al. (2020). Solid lines: results from the present calculations with the fiducial Galactic model.}
\label{fig-compar-long}
\end{figure}
 \begin{figure}[h!]
 \center{
\includegraphics*[height=5.8cm,width=6.cm]{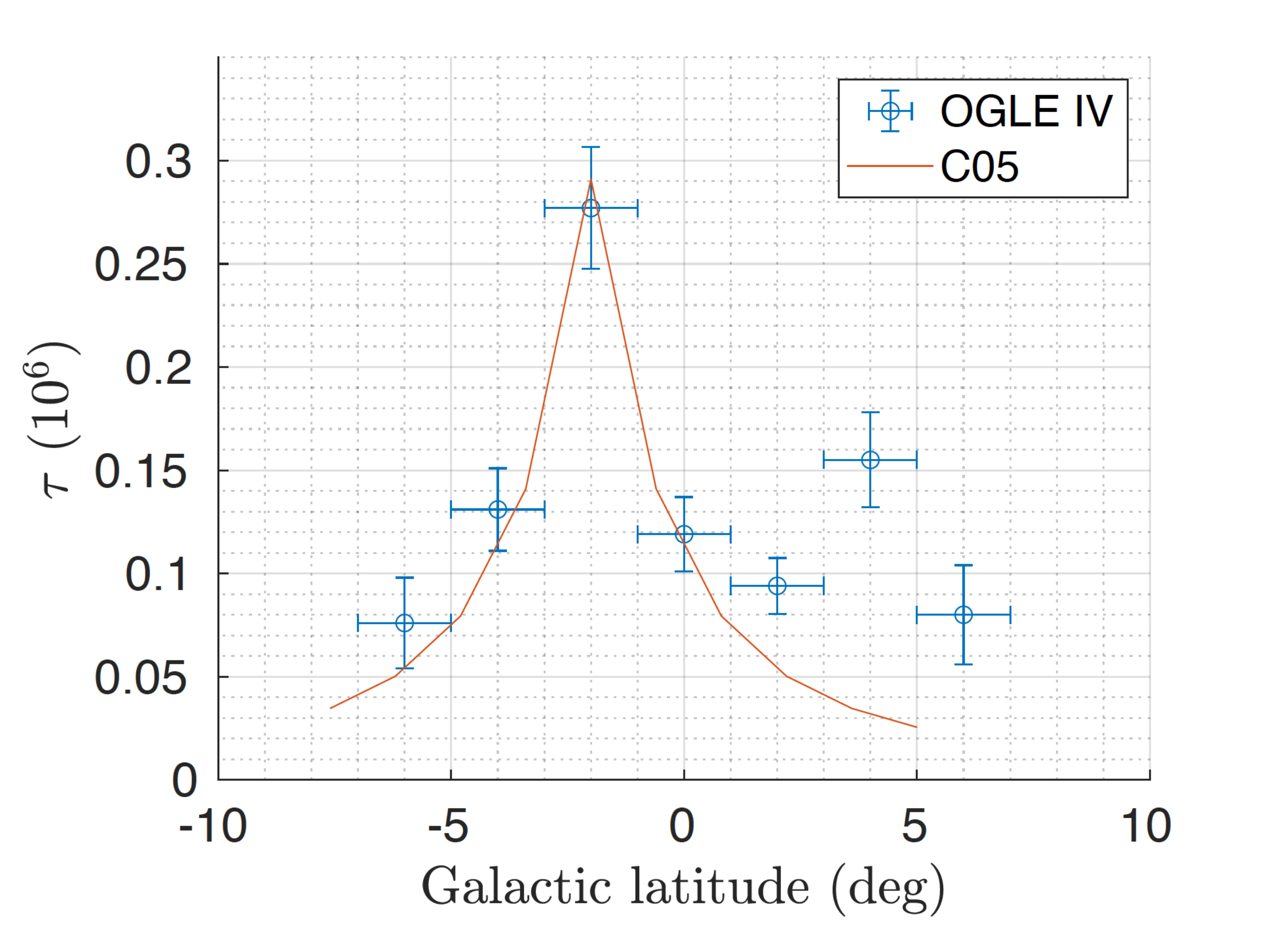}
\includegraphics*[height=5.8cm,width=6.cm]{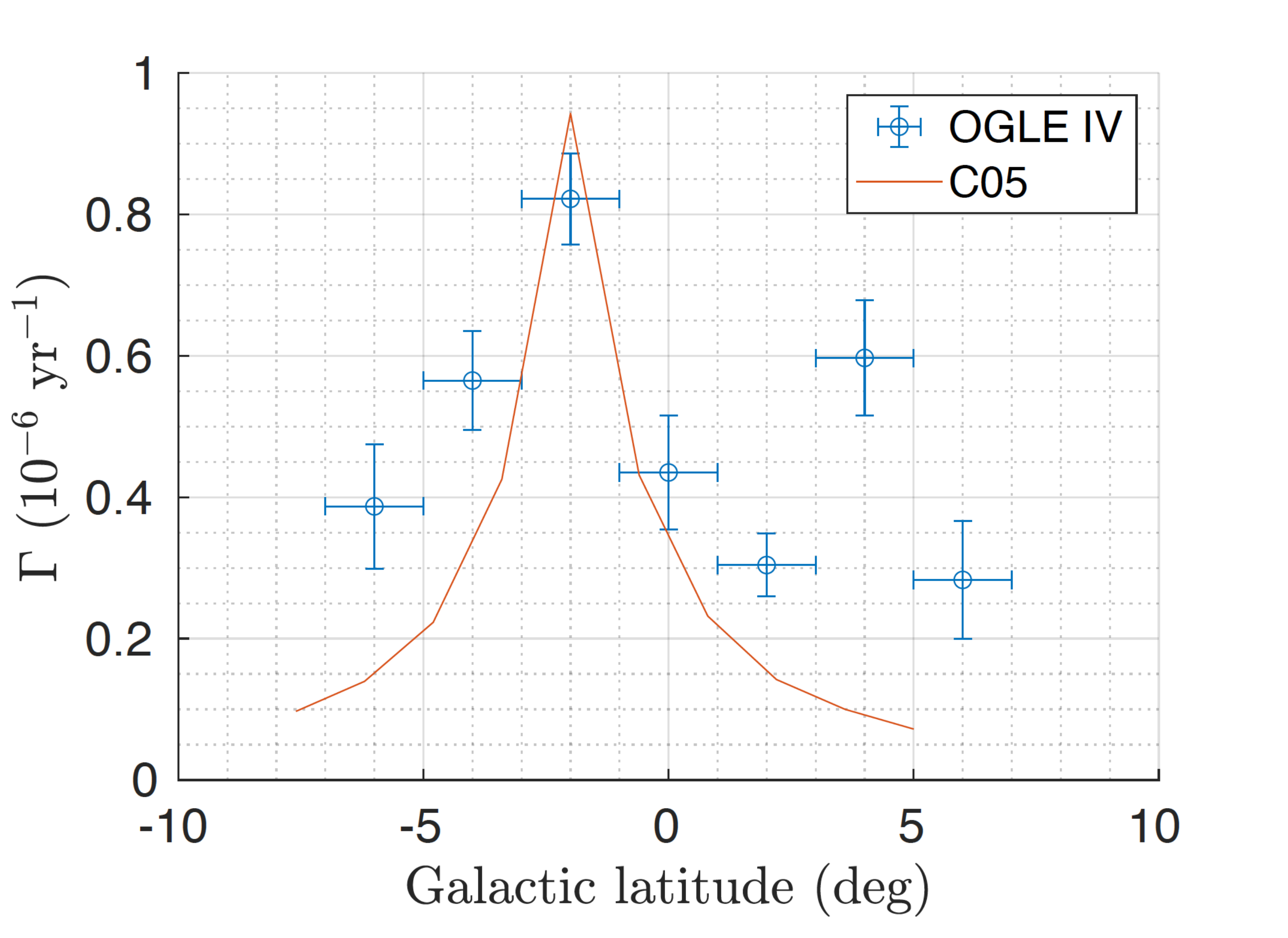}
\vspace{-0.3cm}
}
  \caption{Same as Figure \ref{fig-compar-long} as a function of latitude.}
\label{fig-compar-lat}
\vspace{.5cm}
\end{figure}

The agreement between the calculations and the data is quite satisfactory  for the longitude dependence but is not as good for the latitude, notably at high latitude. As seen in the figure and shown in Mr\'oz et al. (2019, 2020), the optical depth and event rate exponentially decrease with the angular
distance from the Galactic Center. In both cases, we note the correlation between $\tau$ and $\Gamma$. The fact that both the $\tau$ and $\Gamma$ distributions peak at a latitude $b=-2^0$ most likely reflects the position of the Sun above the Galactic plane (see \S4.2.2).
This is in good agreement with the expectations of the Galactic models (Sajadian \& Poleski 2019). The strong deviation for the central fields is not surprising, as the integrated density varies substantially as one approaches the Galactic center. A more precise analysis would require a much better sampling than the one done by Mr\'oz et al. (2020). Indeed, the authors focussed their study on the fields in
the plane and not toward the bulge (see \S\ref{central}).  We have no clear explanation for the disagreement
between the model and the data for the high latitudes. A plausible explanation would be overdensities at these locations, yielding an excess of lenses originating from the enhanced stellar density in these regions, with sources in the background disk. Such an excess of microlensing events was recently detected with the Gaia Data Release 3 (Wyrzykowski et al. 2022) that coincides with the Gould Belt. This is of course just a suggestion.

Overall, our fiducial model, based on the C05 IMF, is globally consistent with the observations toward the Galactic plane. A more refined sampling could probably help improve the
model, notably the density distribution as a function of $(b,l)$.

\subsection{ Summary of the Comparisons with the Observations. Consequences for Star and Brown Dwarf Formation}

All of these comparisons show that the C05 IMF is consistent with the recent OGLE-IV constraints over the entire distributions, except 
 for events $\lesssim 20$ days for the OGLE-IV central-most fields. In \S\ref{central}, we suggest possible explanations for this behavior, including a bottom-heavy IMF in the central parts of the Galactic bulge.
Although  it might be worth performing the same type of detailed numerical simulations as in Wegg et al. (2017) in order to optimize the $(m_0,\sigma)$ parameters in the C05 IMF (see also Equation (34) of Chabrier et al. 2014a),
the present results suggest that the optimized parameters should differ only modestly from the latter, given the proximity of
the theoretical and experimental distributions. In
contrast, even though one must remain cautious about potential biases in the microlensing experimental analysis, the C03, K01 and A16 IMFs (we recall that the latter yields results very similar to the former  two) seem to be excluded. Because, as explored in Appendix C, the results depend only modestly on the Galactic model, this general conclusion can be considered as reasonably robust.

These results raise an important issue concerning the formation of brown dwarfs. In order to  be consistent with the observed BD 
distributions in young clusters,
calculations based on the K01 IMF need to invoke a strong discontinuity near the star-BD transition  (Thies \& Kroupa 2007), which corresponds to an Einstein time around $\te \sim$ 6 days for the bulge conditions (Equation (\ref{teobs})). 
Based on this result and supposedly a problem with the properties of BD companions (see Chabrier et al. 2014b for a discussion on this issue), these authors argued that BDs have a different IMF from stars and thus form differently. 
We suggest that a more plausible explanation is that the mass probability law
represented by the K01 IMF is incorrect. This is clear from Figure \ref{fig-compar}, which shows that the K01 IMF significantly overestimates the number of events below $\sim 10$-20 days, i.e. the low-mass star to very low mass star domain. Retrieving a correct $\te$-distribution for these events
requires switching from K01 to C05 near the bottom of the main sequence, which indeed implies a discontinuous IMF. As explained in \S\ref{sec-IMF} and Chabrier (2005), this stems essentially from the erroneous normalization of the K01 IMF at the stellar-substellar boundary, based on an obsolete luminosity function. As seen in Figure \ref{fig-IMFs} of App. B and Figure 3 of Chabrier (2005), the Kroupa (2001) IMF predicts more than twice as many low-mass stars at 0.1 $\msol$ than the C05 IMF, 
and globally overestimates the number of LMS below about $\lesssim 0.5\,\msol$. In contrast, the fact that the C05 IMF, which extends continuously over the entire stellar plus substellar
domain, adequately reproduces cluster and field BD distributions, plus the microlensing observations basically down to 
the star formation limit, strongly suggests a common {\it dominant} formation mechanism for stars and BDs (see Chabrier et al. 2014b for a review). This implies that alternative mechanisms, such for instance as disk instability or dynamical ejection, are not the main drivers of BD formation, even though they may contribute more modestly to the process.
The recent results of the WISE survey, for
instance (Kirkpatrick et al. 2019, 2021) suggest a rising number of field very low mass BDs, more consistent with a rising power-law IMF than with a lognormal one, even though the disagreement with the latter remains  within the observational  error bars. 
It should be kept in mind, however, that these determinations imply model-dependent mass-effective temperature transformations and are subject to large uncertainties. Indeed, no atmosphere model can presently be
considered as reliable enough to accurately describe  the spectral energy distribution of these cool, atmospherically complex objects. Interestingly, microlensing experiments, although subject to other limitations, are exempt from such fragile photometric or spectroscopic transformations. 

Similarly, the detection of a rich population of free floating 'planets' in the Upper Scorpus young stellar association has recently been claimed in the literature (Miret-Roig et al. 2022). {It is important to stress that these authors used the IAU definition for planets vs BDs, with a cut-off mass at 10 $\mjup$. As examined in detail in Chabrier et al. (2014b and references therein), this {\it semantic} definition has no robust scientific justification and brings a lot of confusion. The observations of Miret-Roig et al. (2022), if confirmed, rather suggest an excess of low-mass BDs compared with the C05 IMF.
These determinations, however, are subject to both observational and theoretical uncertainties and must be confirmed unambiguously.
It must also be kept in mind that the IMF is not "carved in stone" and can potentially exhibit some local variations. Observations of many nearby young clusters or star-forming regions down to a few Jupiter masses show that all observed sequences are consistent with the same ‘underlying’ C05 IMF. A $\sim$10-30\% or so local variation below $\lesssim 10\,\mjup$ around this underlying IMF, consistent with the analysis of Miret-Roig et al. (2022) or Gould et al. (2022, Fig. 9) is not excluded. Whether this excess, if confirmed, is due to an underestimation of the low-mass BD part of the C05 IMF or reflects a population of ejected or wide-orbit BD companions or planets (formed in a disk) remains an open question for now.
We also recall that the non detection of any excess of very short timescale events in the OGLE microlensing experiments excludes a large population of free-floating or wide-orbit Jupiter-mass objects (Mr\'oz et al. 2017, 2019), showing that announcements that have been made in the past were incorrect. Great caution must thus be taken when claiming excess of very low mass objects.

\section{Conclusion} 
\label{sec:conclu}

In this paper, we have compared the optical depths and the event timescale distributions obtained with 4 different IMFs, namely Chabrier (2003, 2005), Kroupa (2001) and Awiphan et al. (2016) with the results obtained with the recent OGLE-IV experiments toward various regions of the Galactic center or plane. These have characterized a total of 8000 events for total exposures $>10^8$ star-yr, allowing an accurate statistical comparison. The C05 IMF 
extending down to essentially the bottom of the BD domain is fully consistent with the OGLE-IV outer field observations. The new optical depth and event rate analysis conducted with the present calculations eases
the tension between the previous measurements and Galactic models.
In contrast, the C03, K01 and A16 IMFs predict a number of short-time, and thus low-mass events larger than the OGLE-IV distributions and fail to reproduce the proper location of the peak of the distribution. 
This failure of the Kroupa IMF to correctly reproduce the mean durations of microlensing events, with a higher contribution of low-mass objects, inducing a deficit of predicted long-duration events, has already been noted by Moniez et al. (2017). The K01 IMF has also been shown to fail
to reproduce the observed distribution of BDs in various young clusters, in contrast to the C05 IMF (Andersen et al. 2008).
Similarly, the disagreement between the observed present distributions and those obtained with the Awaiphan et al. (2016) IMF, which yields a distribution quite similar to the one obtained from the K01 IMF in the low-mass domain
(see Figure \ref{fig-IMFs} of App. B), not mentioning the peculiar behavior of this IMF at large masses, steeper than the Salpeter slope, raises questions about  the accuracy of this IMF.

 The similarity between the local canonical Chabrier (2005) IMF and the one presently inferred from the microlensing observations toward the Galactic center, which extends well into the BD domain,
points to a rather universal star+BD {\it dominant} formation process for
the origin of the IMF. In other words,  
it shows that, under MW-like conditions, this process only weakly depends upon the environment, including on
stellar feedback (see, e.g., Hennebelle et al. 2020). This challenges numerical simulations that suggest that the IMF strongly varies with the properties of the parent cloud. 
It seems that very extreme conditions, as inferred, e.g., for massive early type galaxies, 
characterized by both significantly higher densities {\it and} velocity dispersions, and thus higher external pressures (so, surface densities) than under MW-like conditions, are required to affect the IMF genesis. 

 For the OGLE-IV central fields, the C05 IMF underestimates the number of short-timescale $(\te \lesssim 20$ days) events compared with the observations. Whether or not the disagreement can be explained by either experimental or theoretical limitations (see \S\ref{obs}) remains an open question. Although an underestimation of detection efficiency does affect the short-time event distribution in these regions, resolving this issue would require significant changes in 
these parameters. As mentioned in \ref{math}, the very detailed procedure performed in Mr\'oz et a. (2017, 2019) makes this solution very unlikely. 
The agreement between the microlensing distributions toward the Galactic centermost parts, which display a larger number of short-timescale events than for the other regions, and a Chabrier et al. (2014a) type bottom-heavy IMF suggests that the central part of the Galaxy indeed formed in a burst-like mode, 
providing high density and turbulence, a scenario which has already been suggested on other grounds. 
 Indeed, as mentioned previously, various observations point to a two-step process in the bulge formation, with the existence of an early strong gas rich accretion phase, triggering a burst of star formation more intense close to the plane that far from the plane (Hasselquist et al. 2020), followed by a more secular evolution
(see, e.g., Grieco et al. 2015). 
It is likely that, under the effect of dynamical frictions, such violent accretion events powered highly turbulent motions (e.g., Dekel \& Burkert 2014, Bournaud et al. 2009). As  shown in Chabrier et al. (2014a), this yields an offset of the normalizations of the Larson density-size and velocity-size relations compared with standard (quiescent) GMC conditions, as observed in GMCs
in starbursting galaxies (e.g., Dessauges-Zavadsky et al. 2009). Even though we are aware of the speculative nature of this kind of suggestion, the aforementioned diagnostics, combined with the present IMF analysis, at least lend some support to such a scenario.
It is worth noting that some centermost young stellar disks close to the supermassive BH show a highly top-heavy IMF. But such formation circumstances should be very rare, as they have not affected most of the central cluster.

Our results are relevant in view of the future microlensing plans with the Roman Space Telescope (formerly WFIRST) in the near-IR.
An additional
reason that makes the study of the IMF in the bulge of spiral
galaxies and elliptical galaxies important is the
possibility that these spheroids could potentially contain the
majority of the stellar mass of the universe (see, e.g.,
Fukugita et al. 1998).

\acknowledgments
The authors are very grateful to Przemek Mr\'oz for providing all the data and efficiency tables and for always kindly answering our questions.
We are also deeply endebted to the referee, whose very useful remarks helped improve the final manuscript. We also thank the referee for providing new values for the bulge velocity dispersions.

\references  

Andersen, M., et al. 2008, \apjl, 683, 183  

Awiphan, S., Kerins, E., \& Robin, A. C. 2016, \mnras, 456, 1666  

Bahcall, J. \& Soneira, R., 1980, \apjs, 44, 73  

Barbosa, C., et al. 2020 , \apjs, 247, 46 

Barbuy, B., Chiappini, C. \& Gerhard, O., 2018, \araa, 56, 223

Bastian, N., et al. 2010, \araa, 48, 339  

Bensby T., et al. 2017, \aap, 605, 89

 Bournaud, F., Elmegreen, B. \& Martig, M., 2009, \apjl, 707, L1

Brand, J. \& Blitz, L., 1993, \aap, 275, 67

Brunthaler, A. et al. 2011, Reviews in Modern Astronomy, 23, 105 

Calchi Novati, S. et al., 2008 \aap, 480, 723  


Calamida, A., et al. 2015, \apj, 810, 8  

Cappellari, M., et al. 2012, Nature, 484, 485

Clarkson, W., et al. 2008, \apj, 684, 1110 

Conroy, C. \& van Dokkum, P.G, 2012, \apj, 760, 71  


Chabrier, G., 2003, \pasp, 115, 763   

Chabrier, G., 2005, ASSL,   327, 41

Chabrier, G., Hennebelle, P. \& Charlot, S., 2014a, \apj, 796, 75

Chabrier, G. et al. 2014b, Protostars and Planets VI, University of Arizona Press, 914, 619

Damian, B., et al. 2021, \mnras, 504, 2557

Debattista, V., et al. 2015, \apj , 812, 16

Deka, M., Deb, S. \& Kurbah, K., 2022, \mnras, 514, 3984

Dekel, A. \& Burkert, A., 2014, \mnras, 438, 1870

Dessauges-Zavadsky, M. et al. 2019, NatAs., 3, 1115  

Dwek, E. et al. 1995, \apj, 445, 716  

Fukugita, M., Hogan, C. J. \& Peebles, P. J. E., 1998, \apj, 503, 518  

Glicenstein, J.-F., 2003, Nuclear Physics B Proceedings Supplements, 118, 527

Goldberg, D., \& Wozniak, P., 1998, Acta Astronomica, 48, 19 

Golovich, N. et al. 2022, \apjs, 260, 2 

Gould, A., 2000, \apj, 535, 928 

Gould, A., 2004, \apj, 606, 319


Gould, A., et al. 2022, JKAS, 55, 173

Gravity Collaboration, 2021, AAP, 647, A59 

Grieco, V. et al. 2015, \mnras, 450, 2094

Gu, M. et al. 2022, \apj, 932, 103

Guszejnov, D. et al., 2017, \mnras, 468, 4093




Hasselquist, S. et al. 2020, \apj, 901, 109

Hennebelle, P. et al. 2020, \apj, 904, 194

Henry, T. \& McCarthy, D., 1990 , \apj, 350, 334


Hopkins, P., 2013, \mnras, 433, 170  

Horta D., et al. \mnras, 2021, 500, 1385

Iocco, F. et al. 2011, Journal of Cosmology and Astroparticle Physics, 11, 29

Jorgens, V., 2008, \aap,  492, 545

Juri\'c, M.,  et al. 2008, \apj, 673, 864

Kiraga, M. \& Paczy\'nski, B., 1994, \apjl, 430, L101  

Kirkpatrick,  D. et al. \apj, 2019, 240, 19

Kirkpatrick,  D. et al. \apj, 2021, 253, 7

Kroupa, P., 2001, \mnras, 322, 231

Kunder, A., et al. 2020, \aj, 159, 270

Launhardt, R., Zylka, R. \& Mezger, P. G., 2002, \aap, 384, 112  

Lian, J., et al. 2020a, \mnras, 497, 2371  

Lian, J., et al. 2020b, \mnras, 497, 3557  

McWilliam, A. \& Rich, R.M.,  1994, \apjs, 91, 749

Majaess, D. J., et al. 2009, \mnras, 398, 263 

Maraston, C., 1998, \mnras, 300, 872

M\'era, D., Chabrier, G. \& Schaeffer, R., 1998, \aap, 330, 937

Miret-Roig, N. et al. 2022, Nat. As., 6, 89  

Moniez, M.,  2010, GReGr., 42, 2047

Moniez, M., et al. 2017, \aap, 604, 124

Mr\'oz, P. et al. 2017, \nat, 548, 183  

Mr\'oz, P. et al. 2019, \apjs, 244, 29  

Mr\'oz, P. et al. 2020, \apjs, 249, 16  


Nataf, D.M. et al. 2013, \apj, 769, 88

Navarro, M-G. et al. 2017, \apjl, 851, 13
 
Nishiyama, S. et et al. 2013, \apjl, 769, 28  

Paczynski, B., 1996, \apj, 304, 1

Parravano, A. et al. 2011, \apj, 726, 27  

Pasetto, S., et al. 2012a, \aap, 547, 70  

Pasetto, S., et al. 2012b, \aap, 547, 71 

Peale, S., 1998,  \apj, 509, 177  

Popowski, P. et al. 2005, \apj, 631, 879  

Portail, M., et al.  2015, \mnras, 450, 66

Portail, M., et al. 2017a, \mnras, 465, 1621 

Portail, M., et al. 2017b, \mnras, 470, 1233   

Queiroz, A. et al. 2021, \aap, 

Rahal, Y., et al. 2009, \aap, 500, 1027  

Reid, I.N., et al. 2002, \aj, 124, 2721  

Renzini, A., et al. 2018, \apj, 863, 16

Sahu, K., et al. 2022, \apj, 933, 83

Sajadian, S. \& Poleski, R., 2019, \apj, 871, 205 

Salpeter, E., 1955, \apj,  121, 161 


Savino, A., et al. 2020, \aap, 641, 96

Sch\"onrich, R. et al. 2015, \apjl, 8712, 21  

Sharples, R. et al. 1990, \mnras, 246, 54  

Smith, R., 2020, \araa, 58, 577

Stanek, K.Z., 1995, \apl, 441, 29

Stanek, K.Z. et al. 1997, \apj, 477, 163

Sumi, T., Bennett, D. P., Bond, I. A., et al. 2013, \apj, 778, 150  

Sumi, T., Kamiya, K., Bennett, D. P., et al. 2011, \nat, 473, 349\

Sumi, T., \& Penny, M. T. 2016, \apj, 827, 139   

Thies, I. \& Kroupa, P., 2007, \apj, 671, 767  

Treu, T., et al., 2010, \apj, 709, 1195  

Udalski, A., Szymański, M. K. \& Szymański, G., 2015, Acta Astronomica, 65, 1

van Dokkum, P.G. \& Conroy, C., 2012, \apj, 760, 70 

van Dokkum, P.G., 2008, \apj, 674, 29

Wegg, C. \& Gerhard, O.,  2013, \mnras, 435, 1874

Wegg, C., et al. 2015, \mnras, 450, 4050

Wegg, C., Gerhard, O., \& Portail, M. 2017, \apjl, 843, L5  

Wood, A., 2007, \mnras, 380, 901



Wyrzykowski, L. et al., 2015, \apjs, 216, 12  

Zoccali, M., 2019, BAAA, 61

Zhao, H., 1996, \mnras, 283, 149

Zhao, H. \& Mao, S., 1996, \mnras, 283, 1197

Zheng, Z., et al., \apj, 2001, 555, 393
  

\onecolumngrid
\appendix

\section{A. Microlensing calculations} \label{ap:micro}

The microlensing calculations are the same as described in Appendix A of M\'era et al. (1998, MCS98), except for the density of the deflectors (see \ref{sec_lens}) and are summarized here.

\subsection {A1. Characteristic time:}

The duration $t$ of a microlensing event, defined as the time during
which the magnification of the monitored star is larger than a given
threshold amplification $A_T$ (usually
$A_T=1.34$), reads:

\begin{eqnarray} 
t=\frac{2\re}{v_\perp}\sqrt{u_T^2-u_{min}^2}\simeq 45\
\mathrm{days}\ \frac{100\, \mathrm{km.s}^{-1}}{v_\perp}\times
\sqrt{\frac{M}{0.1\msol}} \sqrt{\frac{D_S}{10\, \mathrm{kpc}}}
\frac{\sqrt{x(1-x)}}{0.5}\sqrt{u_T^2-u_{min}^2}
\label{t}
\end{eqnarray}

\noindent where $v_\perp$ is the lens transverse velocity w.r.t. the line
of sight, $u_T$ (resp. $u_{min}=d_{min}/\re$) is the dimensionless impact parameter
corresponding to the threshold amplification (resp. maximum
amplification) $A_T$ (resp. impact parameter $u=1$ at $d=\re$), $D_S$ is the distance to the source, $D_L=xD_S$ and $M$
denote, respectively, the distance and the mass of the lens, and
$\re=\frac{2}{c} \sqrt{GMD_Sx(1-x)} \simeq 1.4\,{\rm AU}\,\sqrt{\frac{M}{0.1\msol}} \sqrt{\frac{D_S}{10\, \mathrm{kpc}}} \frac{\sqrt{x(1-x)}}{0.5}$ is the Einstein radius, with ($\pi \re^2$) the surface of the Einstein disk.
Note that, given the fact that $v_\parallel /{D_S}\simeq 110\
\mathrm{days}\times 100 \mathrm{km.s}^{-1} / 10\, \mathrm{kpc} \approx 10^{-9}\ll 1$ and $x\simeq 1$, the variation along $x$ is negligible and $t$ only depends on $v_\perp$.

The {\it characteristic} time of an event is defined as:

\begin{eqnarray} 
\te \equiv \te(M,x,v_\perp) &=& \frac{\re}{v_\perp}  = \frac{t}{2\sqrt{u_T^2-u_{min}^2}} = \frac{2}{cv_\bot}\sqrt{GD_SMx(1-x)} \\ 
&\simeq&25\, \mathrm{days}\ \frac{100\, \mathrm{km.s}^{-1}}{v_\perp}\times
\sqrt{\frac{M}{0.1\,\msol}} \sqrt{\frac{D_S}{10\, \mathrm{kpc}}}
\frac{\sqrt{x(1-x)}}{0.5}.
\label{te}
\end{eqnarray}
Using more observable quantities, this equation can be rewritten as

\begin{eqnarray} 
\te &=& \frac{\theta_E}{\mu_{\rm rel}}  = \frac{\sqrt{\kappa M \pi_{\rm rel}}} {\mu_{\rm rel}} \simeq 6.4\,{\rm days}\,\frac{6.5\,{\rm mas/yr}}{\mu_{\rm rel}}\sqrt{\frac{M}{0.1\msol}} \sqrt{{\pi_{\rm rel}}\over{0.016\,{\rm mas}}}, \label{teobs}\\
\theta_E &=& \sqrt{\kappa M \pi_{\rm rel}} = 0.32\,{\rm mas}\, ({1-x\over x})^{1/2} ({D_S\over 8\,{\rm kpc}})^{-1/2} ({M\over 0.1 \,\msol})^{1/2},  
\end{eqnarray}
where $\theta_E$=$R_E/D_L$ is the Einstein angular radius, $\kappa={4v_\oplus^2\over \msol c^2}={4G\over c^2}\,{\rm AU}^{-1}=8.144$ mas $\msol^{-1}$, with $v_\oplus\sim$ 30 $\kms$ the speed of Earth, $\pi_{\rm rel}=\pi_l-\pi_s={{\rm AU}\over D_L^{-1}-D_S^{-1}}$ is the lens-source relative parallax, and $\mu_{\rm rel}=\mu_l-\mu_s=\mu_E\theta_E = {v_\perp \over D_L}$ is the relative lens-source proper motion. The values $\mu_{\rm rel}=6.5$ mas/yr and $\pi_{\rm rel}\simeq0.016$ mas in equation (\ref{teobs}) correspond to the typical values  for bulge-bulge lensing.

The event timescale distribution is given by $P(\te)$=$\int P(M,x,v_\perp) \delta(\te-\frac{\re}{v_\perp}) dM\,dx\,dv_\perp$ with $P(M,x,v_\perp)=P_{{eff}_M}(M) P_{{eff}_x}(x) P_{{eff}_v}(v_\perp)$, where $P_{{eff}_M}(M), P_{{eff}_x}(x), P_{{eff}_v}(v_\perp)$ denote the {\it effective} probability distributions calculated below (eqns.(\ref{Pm})-(\ref{Pv})).

This yields the average characteristic time $\langle  \te \rangle$:

\begin{eqnarray}
\langle  \te\rangle=\frac{2\sqrt{GD_S}}{c}\langle  \sqrt{M}\rangle\langle  \frac{1}{v_\bot}\rangle\langle  \sqrt{x(1-x)},
\label{temoy}
\end{eqnarray}
where the means are calculated with the aforementioned effective probabilities.

\subsection {A2. Lens density}
\label{sec_lens}

Assuming for now that all of the sources are located at the distance $D_S$, the number of lenses located at $D_L$ with a mass $M\in [M,M+dM]$ is
$dN_{\rm lens}=n_{\rm lens}(D_L,M)\times 4\pi (D_L)^2 d(D_L)dM$. 
The density depends on the probability density $P(M)$ for a lens to have a mass $M\in [M,M+dM]$, i.e., of the mass function, $\xi(M)$.
The number density of lenses at a distance $D_L$ thus reads (e.g., De Rujula et al. 1991)

\begin{eqnarray}
    n_{\rm lens}(D_L,M)=\frac{\rho(D_L)}{\langle M \rangle}\,\xi(M), \label{eqn_nlens}
\end{eqnarray}
where $\rho(D_L)$ is the Galactic mass density at $D_L$, so $\frac{\rho(D_L)}{\langle M \rangle}$ corresponds to the number density of starlike objects, i.e., of potential deflectors at $D_L$. 

\noindent This differs from the expression used in MCS98, which  is $n_{lens}(D_L,M)=\frac{\rho(D_L)}{ M }\,\xi(M)$. Even though the two expressions look similar, they yield different results. Indeed, given the fact that the Einstein radius is proportional to the square root of the mass, $\re \propto \sqrt{M}$, the mass dependence of the rate, $\Gamma$, is $\Gamma\propto 1/\sqrt{M}$ in the former case, while it is $\Gamma\propto \sqrt{M}$ with Equation(\ref{eqn_nlens}). This means that the distribution of events favors high masses in the former case, low masses in the second one. We verified that using the MCS98 normalization yields a fraction of short time events much larger than the one observed by OGLE, in contrast to Equation(\ref{eqn_nlens}).

\subsection {A3. Optical depth}
\label{sec_tau}

From the definition of a microlensing event, a lens $(x,M,t)$ covers a solid angle
\begin{eqnarray} 
\delta \Omega={ u_\perp(\pi \re)^2\over 4\pi(D_L)^2}= \frac{GM}{c^2}\frac{(1-x)}{D_L}.
\end{eqnarray}
The number of lensed objects at a distance $D_L$ is  $\frac{\rho(D_L)}{\langle M \rangle}\times (\pi \re^2)$.
The optical depth up to a distance $D_S$, i.e., the probability for a source star located at $D_S$ to be microlensed by a factor $\ge A_T$ at a given time is given by:

\begin{eqnarray} 
\tau(D_S)&=&\int \delta \Omega \int_0^\infty  \, n_{\rm lens}(D_L,M)\times 4\pi (D_L)^2 d(D_L)dM\\
&=& \int_0^1 u_T^2 \pi \re^2 \frac{\rho(D_L)}{{\langle M \rangle}} D_S dx\\
&=&4\pi \frac{GD_S^2}{c^2} \int_0^1     \rho(D_L) x(1-x) dx,
\label{tau}
\end{eqnarray}
where we have taken $u_T=1$. Note that the optical depth does not depend  on either the mass or the velocity of the deflector.

\noindent For observations, for a population of $N_s$ source stars observed during a total duration $T_{obs}$, the probability that a star is amplified corresponds to the sum of all the observed event duration divided by the exposure
$E=N_s\times T_{obs}$. The {\it experimental} optical depth measured by the observations is thus:

\begin{eqnarray}
\tau_{exp}={\pi \over 2}\frac{1}{E}\sum_i \frac{t_{i,obs}}{\epsilon(t_i)},
\label{tauexp}
\end{eqnarray}
where  $\epsilon(t)$ denotes the detection efficiency,
 $t_{i,obs}$ is the observed Einstein radius crossing time of the $i$-th event and $\epsilon(t_i)$ is the detection efficiency at this timescale.

\subsection {A4. Event Rate:}

During a timescale $dt$, the  surface covered by a lens per unit time is $\delta S=2u_T\re v_\bot dt$, which corresponds to a solid angle $\delta \Omega = \delta S/(D_L)^2$. For a number of observed stars $N_s$, the
number of events is thus $dN=N_s\, \delta S\, n_{lens}(D_L,M) P(v_\bot)\,dM\, dv_\bot \, d(D_L)\,dt$, so the event rate for a given Galactic model, i.e., the expected number of events per unit time,  reads:

\begin{eqnarray}
{\rm d\Gamma} = \frac{dN}{N_s dt}=2u_T\re v_\bot \frac{\rho(D_L)}{\langle M \rangle}
\xi(M) P(v_\bot)dM\, dv_\bot \, d(D_L),
\label{dGamma}
\end{eqnarray} 
where P($M$)$\equiv \xi(M)$ and P($v_\perp$) are the probability distributions  of, respectively, lens
mass and velocity.  

 If the velocity distribution is independent of the
position, i.e., for a well-defined part of the Galaxy, the integration of (\ref{dGamma}) yields:

\begin{eqnarray}
\Gamma  =  u_T\frac{4\sqrt{GD_S}}{c}
\int_{\minf}^{\msup} \frac{\sqrt{M}}{{\langle M \rangle}} \xi (M) \,dM
\times  \int_0^1 \sqrt{x(1-x)} \rho(D_L) D_S \,d x \int_0^\infty dv_\bot v_\bot P(v_\bot).
\label{Gamma}
\end{eqnarray}
Equation (\ref{Gamma}) shows that the {\it effective} microlensing
probability distributions for the variables $M,x$ and $v_\perp$  are different  from their intrinsic probabilities, because of the dependency induced by the threshold condition $u_T\ge d/\re$. These effective probabilities are given by :

\begin{eqnarray}
&P_{{eff}_M}(M)& \propto \frac{\sqrt{M}}{{\langle M \rangle}} \xi (m) \label{Pm}\\
 &P_{{eff}_x}(x)& \propto \sqrt{x(1-x)} \rho(D_L) \label{Px}\\
 & P_{{eff}_v}(v_\perp)& \propto v_\bot P(v_\bot) \label{Pv} .
  \end{eqnarray}
Note the difference in the effective mass probability with Equation(A5) of MCS98. The observational event rate is determined as:
\begin{eqnarray}
\Gamma_{exp}=\frac{1}{E}\sum_i \frac{1}{\epsilon(t_i)}.
\label{gammaexp}
\end{eqnarray}

  It is easy to show that the optical depth, the event rate and the average characteristic time obey the relation  
\begin{eqnarray}
\tau = \frac{\pi}{2} u_T \times \Gamma \times \langle  \te\rangle.
\end{eqnarray}  

A given Galactic model implies a timescale distribution
$d\Gamma/d\te=\Gamma\times P(\te)$ and the number of events predicted
by the theory for an exposure $E$ is given by:

\begin{eqnarray} N_{th}=E\times \int_0^{+\infty} \epsilon (\te)
\frac{\mathrm{d}\Gamma}{\mathrm{d}\te} \mathrm{d}\te,
 \label{NthA}\end{eqnarray}  
  to be compared with the number of observed events $N_{obs}=\Gamma_{exp} \times E $.

In equation (\ref{temoy}), the means are computed with the effective
probability distributions (\ref{Pm}-\ref{Pv}). When explicitely writing
 the corresponding integrals, we can identify the optical
depth and the event rate, which yields

\begin{eqnarray}\tau = \frac{\pi}{2}u_T\times\Gamma\times\langle  \te\rangle.\label{tau_te_rel}\end{eqnarray}

The numerical calculation of the event rate is detailed below.

\subsection{A5. Distance of the Source Star}

In the above calculations, the distance $D_S$ to the source star is taken to be constant. For the observations toward the bulge, $D_S$ varies because of
the elongation along the line of sight, which implies an extra integral on $D_S$. Denoting $\nu_s$ as the density of source stars {\it visible}
at the distance $D_S$, the total number of stars we can detect between $D_S$ and $D_S+dD_S$ is $N_s=\int_0^\infty \nu_s(D_S) D_S^2 dD_S$. 
This introduces a new probability density for the distance to the source star, $P(D_S)\propto D_S^2\nu_s(D_S)$.
The optical depth up to a given source distance $L$ for a given stellar population thus now reads (Moniez 2010, Zhao \& Mao 1996, MCS98):

\begin{eqnarray}
\tau (L)= \int_0^{L} D_S^2 \nu_s(D_S)  \int_0^1 u_T^2 \pi \frac{4GD_S^2}{c^2} \rho(D_L) x(1-x) dxd D_S. 
\label{tau_dvar}
\end{eqnarray}
The {\it averaged} optical depth over the line of sight for a given stellar population is defined as:

\begin{eqnarray}
\langle \tau \rangle=\frac{ \int_0^{\infty} D_S^2 \nu_s(D_S) \int_0^1 u_T^2 \pi \frac{4GD_S^2}{c^2} \rho(D_L) x(1-x) dxd D_S}
{\int_0^\infty D_S^2 \nu_s(D_S)d D_S}
\label{tau_dvar_mean}
\end{eqnarray}
Similarly, the event rate (\ref{Gamma}) now reads:

\begin{eqnarray} 
\Gamma = \frac{4\sqrt{G}}{c\int_0^\infty \nu_s(D_S) D_S^2 d D_S}
\int_{\minf}^{\msup} \sqrt M \frac{\xi(m)}{\langle M \rangle} d M 
\int_0^{\infty} \nu_s(D_S) D_S^{3.5} d D_S  \nonumber \\
\times \int_0^1 \rho(D_L) \sqrt{x(1-x)} dx 
\int_0^\infty v_\perp P(v_\perp) d v_\perp,  
\label{Gamma_dvar}
\end{eqnarray}
where we have taken $u_T=1$.

\subsection{A6. Density of  Source Stars}
\label{source_stars}

\noindent The determination of $\nu_s(D_S)$, the density of source stars {\it visible} at a distance $D_S$, which enters Equations (\ref{tau_dvar})-(\ref{Gamma_dvar}), requires a luminosity function for these stars. We have used the one derived by Kiraga \& Paczy\'nski (1994), where the number of stars brighter than some absolute luminosity $A_{lim}$ is proportional to $A^\beta$, $N(A>A_{lim})\propto A_{lim}^\beta$, with $-3\le \beta \le -1$. Since the luminosity $A\propto D_S^2$, this yields:

\begin{eqnarray} 
\nu_s(D_S) \propto \rho_s(D_S) D_S^{2\beta+2}.
\label{nu_s}
\end{eqnarray}
The value of $\beta$ depends on the stellar population. Red Clump Giants are bright enough to be observed throughout the bulge, so their luminosity does not depend on the distance, i.e., $\beta=0$.
For the other source stars located in the Galactic center, we have taken $\beta=-1.2$ for our best fiducial model (see \S\ref{OGLE-all}). 
The simple power-law parameterization (\ref{nu_s}) is probably too simplistic to adequately reproduce the variations
of the complete luminosity function from bright to faint stars (Stanek 1995, Wood 2007). It is, however, a reasonable representation
of the latter at faint magnitudes (magnitude $I\gtrsim 16$), where bulge dwarfs dominate the sample and contribute dominantly to the optical depth  (Wegg et al. 2016).
The impact of the parameter $\beta$ upon $\tau$, $\Gamma$, $\langle \te \rangle$ and the timescale distribution is examined in \S\ref{tau-gamma} and \S\ref{OGLE-all} and discussed in more detail in \S\ref{source}.
\\

The integral (\ref{Gamma_dvar}), with six non-independent variables ($M,x,D_S,v_{lens},v_{source},v_\odot$), is calculated with a Monte Carlo integration method (see MCS98 App. A48). Each simulation was carried out with $10^7$ points for each field.
 The limit of the integral  of the optical depth (\ref{tau_dvar_mean}) and the event rate (\ref{Gamma_dvar}) for the distance of the source were chosen as $(D_{min},D_{max})=(0.8,20)$ kpc. Extending these limits is inconsequential. Indeed, the density for $D< 800$ pc is very small compared with the one of the bulge (Kiraga \& Paczy\'nski 1994, Peale 1998) and, given the exponentially decreasing disk and bulge densities, the results become essentially insensitive to   $D_{max}$ beyond this limit (less than $0.1\%$ variation on $\tau$). 
Taking into account the experimental efficiency is  straightforward with a rejection algorithm.

\section{B. Characteristics of the mass functions}
\label{massfunc}

Table \ref{tab-IMF}  summarizes the parameters of the 4 IMFs for {\it resolved objects}, which is relevant for microlensing experiments, compared in the present calculations. The difference between the K01, C03 and C05 IMFs,
normalized to the Hipparcos determination, is shown in Fig 3 of Chabrier (2005).
Note that a slightly modified form of the C05 IMF has been proposed in Chabrier et al. (2014a, Eq.(34)), which ensures not only continuity of the function but also of its first derivative, as suggested by van Dokkum (2008).
These IMFs are portrayed in Figure \ref{fig-IMFs}.

\begin{table*}[!t]
    \centering
    \caption{Parameters of the initial mass function, $\xi(\log m)=dN/d\log \,m$.}
    \begin{tabular}{c c c c}
    \hline
    \hline
\small      Kroupa (2001) & Awiphan (2016) &  Chabrier (2003) & Chabrier (2005)   \\
       \hline
       \hline
        $\xi(\log m)=A_i m^{1-\alpha_i}$                   &  $\xi(\log m)=A_i m^{1-\alpha_i}$  & \\
              \hline
\footnotesize    0.01$\le{m}\le 0.08$: $\alpha_1=0.3$ & \footnotesize $\alpha_1=0.4$ & &     \\
\footnotesize          ($A_1$=4.575)   & \footnotesize $A_1$=4.034\,(thin d.), 2.708\,(thick d.), 3.762\,(bulge)    &    &  \\ 
\footnotesize   0.08 $\le{m} \le 0.5$: \,\,\,\,$\alpha_2=1.3$ &  &  &       \\
\footnotesize            ($A_2$=0.366)   &      &    &  \\
\footnotesize      $m \ge 0.5$\,\,\,\,\,\,\,\,\,\,\,\,\,\,: $\alpha=2.3$   &   & & \\
              \footnotesize  ($A_3$=0.183)   &     &    &  \\ 
               & {\bf thin disk:} & & \\
   & \footnotesize 0.08 $\le{ m} \le 1.0$: $\alpha_2=1.6$\,\,($A_2$=0.193) &  &       \\ 
    & \footnotesize $1.0 \le {m}$: \,\,\,\,\,\,\,\,\,\,\,\,\,\,\,\,\,$\alpha_3=3.0$\,\,($A_3$=0.193) &  &       \\   
               & {\bf thick disk:} & & \\           
                      & \footnotesize 0.08 $\le{m} \le 0.15$: $\alpha_2=0.5$\,\,($A_2$=2.104) &  &  \\
                      &  \footnotesize 0.15 $\le{m} \le 1.0$: \,   $\alpha_3=1.5$\,\,($A_3$=0.315) &  & \\
                       & \footnotesize $1.0 \le{m}$: \,\,\,\,\,\,\,\,\,\,\,\,\,\,\,\,\,\,\,\,\,$\alpha_4=3.0$\,\,($A_4$=0.315) &  &       \\ 
                 & {\bf bulge:} & & \\  
                      & \footnotesize 0.08 $\le{m} \le 0.70$: $\alpha_2=1.5$\,\,($A_2$=0.236) &  &  \\
                       & \footnotesize $0.7 \le{m}$: \,\,\,\,\,\,\,\,\,\,\,\,\,\,\,\,\,\,\,\,\,\,\,$\alpha_3=2.35$\,\,($A_3$=0.175) &  &       \\ 
       \hline
    \hline
    
                      &   &   ${m}  \le 1$: $\xi(\log m)= A^-\,\exp[-{(\log \, m - \log \, m_c)^2 \over 2\sigma^2}]$    \\ 
                            \hline
                      &        &  \footnotesize $A^-    $ =   $0.642$  & \footnotesize $A^-    $ = 0.7305    \\ 
                      &     &     \footnotesize $m_c$ =   $0.079_{+0.021}^{-0.016}$  & \footnotesize $m_c$ =0.2    \\
                     &                                                           & \footnotesize $\sigma$=   $0.69_{+0.05}^{-0.01}$   & \footnotesize $\sigma$=0.55   \\
                           \hline
                      &         &   ${m} \ge 1$: $\xi(\log m)= A^+ m^{1-\alpha}$ &\\
                            \hline
                     &          &  \footnotesize $A^+    $ =   $0.179$  & \footnotesize  $A^+    $ = 0.326    \\
                      &        &  \footnotesize$\alpha=1.35\pm 0.3$ &  \footnotesize$\alpha=1.35\pm 0.3$ \\
      \hline
    \end{tabular}
\normalsize
\tablecomments{Masses ($m,m_0$) are in $\msol$. The constant $A$ (in $(\log \msol)^{-1}$ pc$^{-3}$) corresponds to IMFs normalized  such that $\int_{0.01\msol}^{100\msol} \xi(m)\, {\rm d}m$=1.
Proper normalizations for the Galactic disk can be found in Chabrier (2005). }
\label{tab-IMF}
\end{table*}


 \begin{table*}[!t]
    \centering
    \caption{Optical depth and event rate for various parameters of the model for the OGLE-IV all-field data (Mr\'oz et al. 2019). All models take a value $\beta=-1.2$ in Equation(\ref{nu_s}).}
    \begin{tabular}{c c c }
    \hline
    \hline
       & $\langle \tau \rangle \,(\times 10^{-6})$ & $\sum \Gamma_{\rm all\,fields}$ ($10^6$ stars/yr)   \\
       \hline
       \hline
OGLE-IV                     &  0.91$\pm 0.13$  & 902 \\
       \hline
Fiducial model              &  0.96  & 928 \\
IMF C03                     & 0.94 & 1007 \\
IMF K01                     & 0.94 & 980  \\
       \hline
        \hline
   {\bf geometry:} &  &     \\
   $\phi=20^o$    &1.08  & 1012 \\
      $\phi=35^o$    &0.88  & 863 \\
   \hline
    {\bf   bulge model:} &  &     \\
   No bulge   & 0.03 & 40 \\
   Stanek '97 & 0.7 & 680 \\
   Zhao '96   & 0.85 & 780 \\
   Deka '22   &  0.97 & 941 \\
   {\bf disk model:} &  &     \\
   No disk   & 0.55 & 601 \\
    Zheng '01 no thick disk & 0.92 & 891 \\
    Zheng '01 with thick disk & 1.1 & 1028 \\
          \hline
 {\bf velocity distribution:} &  &     \\
  $\sigma_{bulge}=110 \kms=\,$constant & 0.96 & 932 \\
    $V_{rot}(R)=220 \kms=\,$constant & 0.96 & 917 \\
   $V_{rot}(R)=$Equation(\ref{vrot}) w/  $\Theta_0=220 \kms$ & 0.96 & 900 \\
    \hline
 {\bf density of source stars:} &  &     \\
 $\rho_s=\rho_{bulge}$  & 0.76  & 720 \\
      \hline
    \end{tabular}
\label{para}
\end{table*}


\begin{figure}[h!]
\vspace{0cm}
\center
\includegraphics[height=12cm,width=12cm]{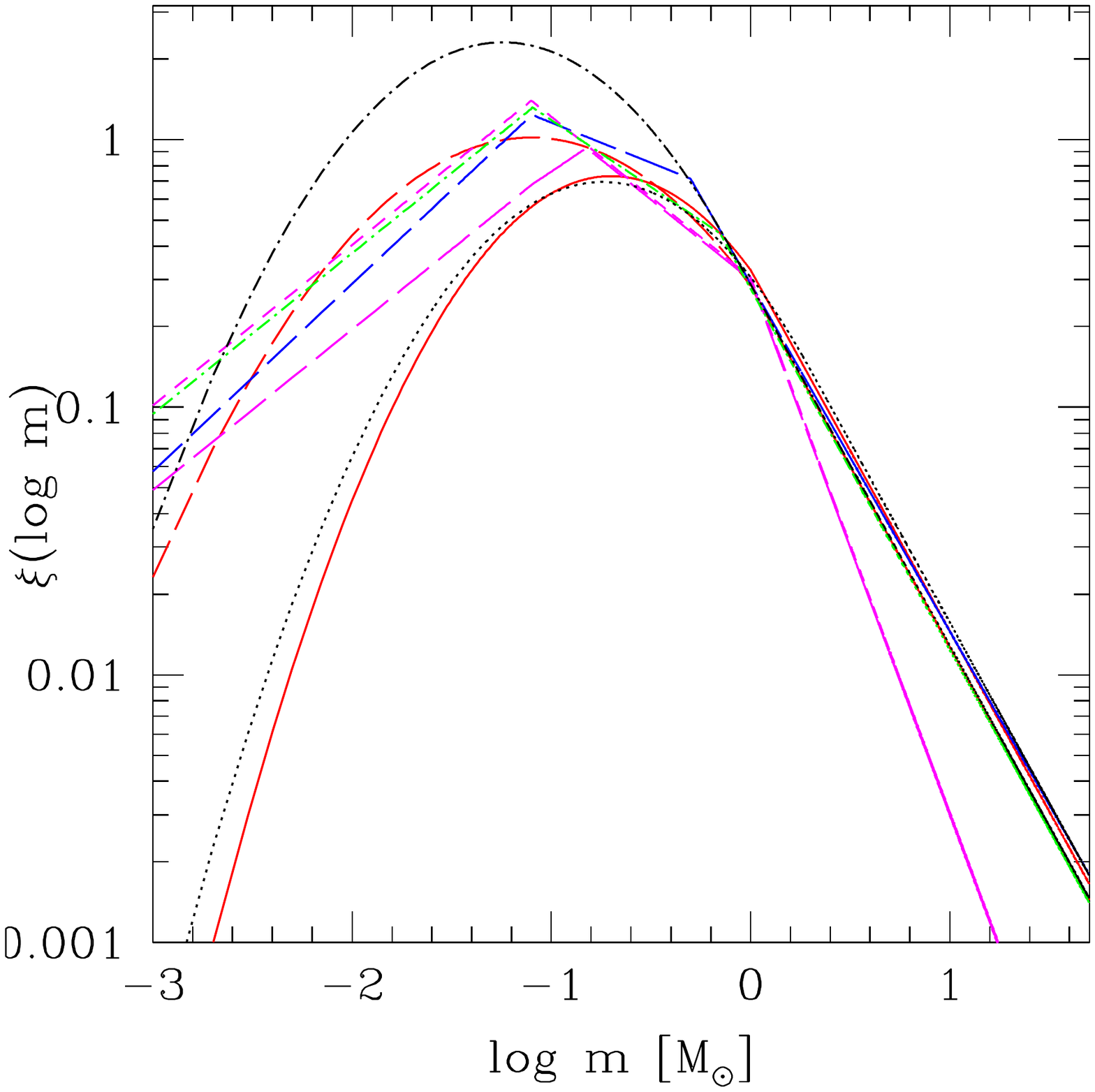}
\vspace{-2.cm}
  \caption{Comparison of  the various IMFs $\xi(\log m)=dn/d\log(m)$ given in Table \ref{tab-IMF}: C05 (solid red), C03 (long-dashed red), K01 (long-dashed blue), A16 thin disk (short-dashed magenta), A16 thick disk (long-dashed magenta), A16 bulge (dashed-dotted green) (note: the values of the slopes of the A16 MFs
  were extended for $m<0.08\,\msol$ by their value for the thick disk, $\alpha=0.5$, to ensure a decreasing IMF in the BD domain). 
  The black dotted and dotted-dashed lines correspond to the functional form given in Equation(34) of Chabrier et al. (2014) which ensures continuity of the derivative. The dotted line reproduces the C05 IMF, for parameters ($m_0$=2.0 $\msol$,  $m_c$=0.182 $\msol$, $\sigma=0.58$, $A_h=0.35$)
  while the dotted-dashed line corresponds to the IMF derived for the OGLE-IV central fields (cf \S\ref{central}) with ($m_0$=0.8 $\msol$,  $m_c$=0.06 $\msol$, $\sigma=0.607$, $A_h=0.09$), both with $\alpha=1.35$.
  The IMFs are normalized  such that $\int_{0.01\msol}^{100\msol} \xi(m)\, {\rm d}m$=1. All the IMFs are normalized at 1 $\msol$ for comparison.}
\label{fig-IMFs}
\end{figure}

\section{C. Dependence  of the results upon model parameters}
\label{depend}

 In this appendix, we examine the impact of various parameters in different models upon the event characteristics, notably the histogram distribution. The efficiency of the fields $\epsilon(t)$ is taken into account.
 The results are displayed in   Figure \ref{fig-param} for the OGLE-IV all fields data (Mr\'oz et al. 2019). Table \ref{para} compares
 the results for the optical depth $\tau$ and the event rate $\Gamma$. All comparisons are made with $\beta=-1.2$ in Equation (\ref{nu_s}).
  
 \begin{figure}[h!]
\vspace{0cm}
\center{
\includegraphics[height=8.5cm,width=8.8cm]{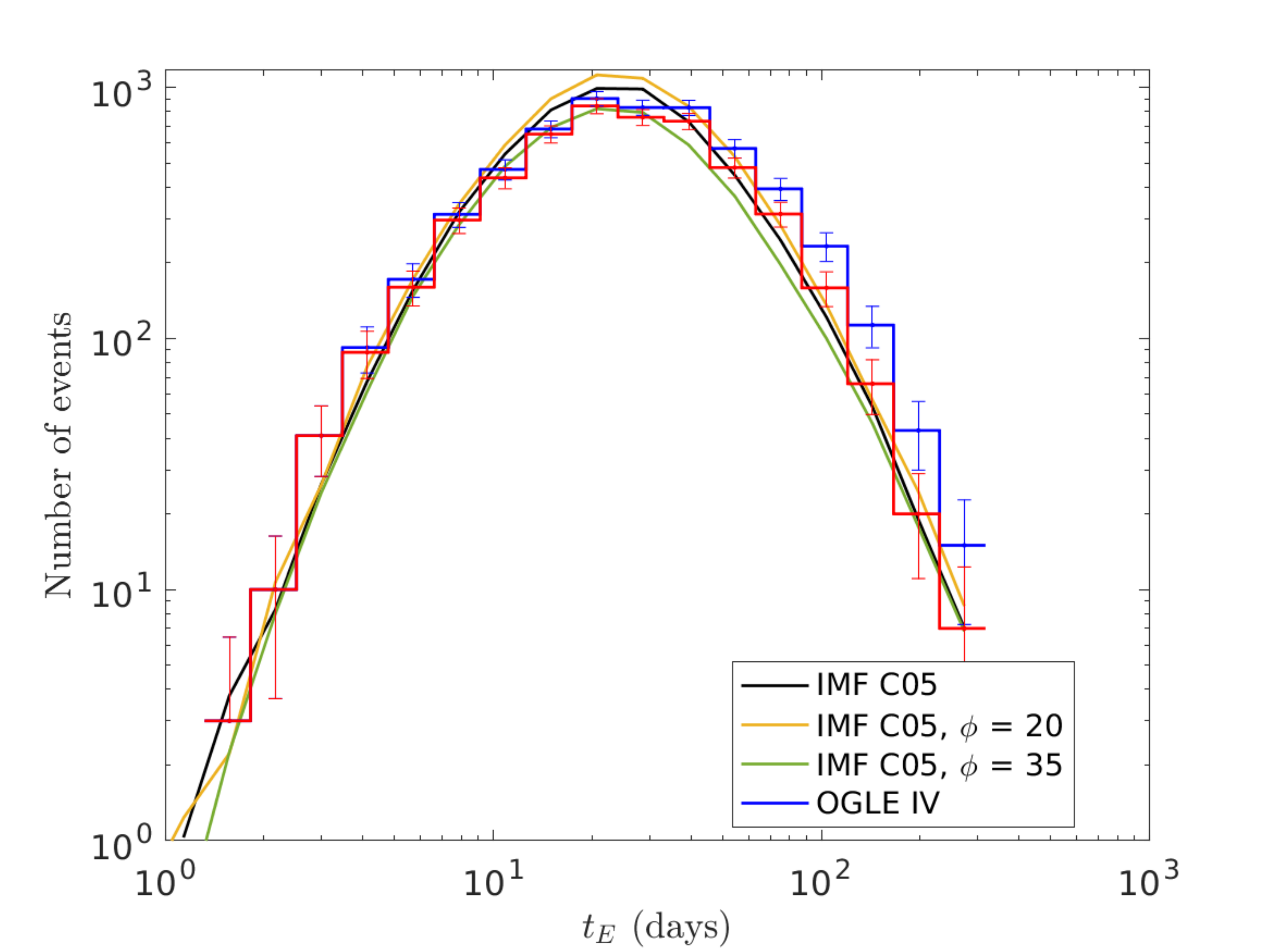} 
\includegraphics[height=8.5cm,width=8.8cm]{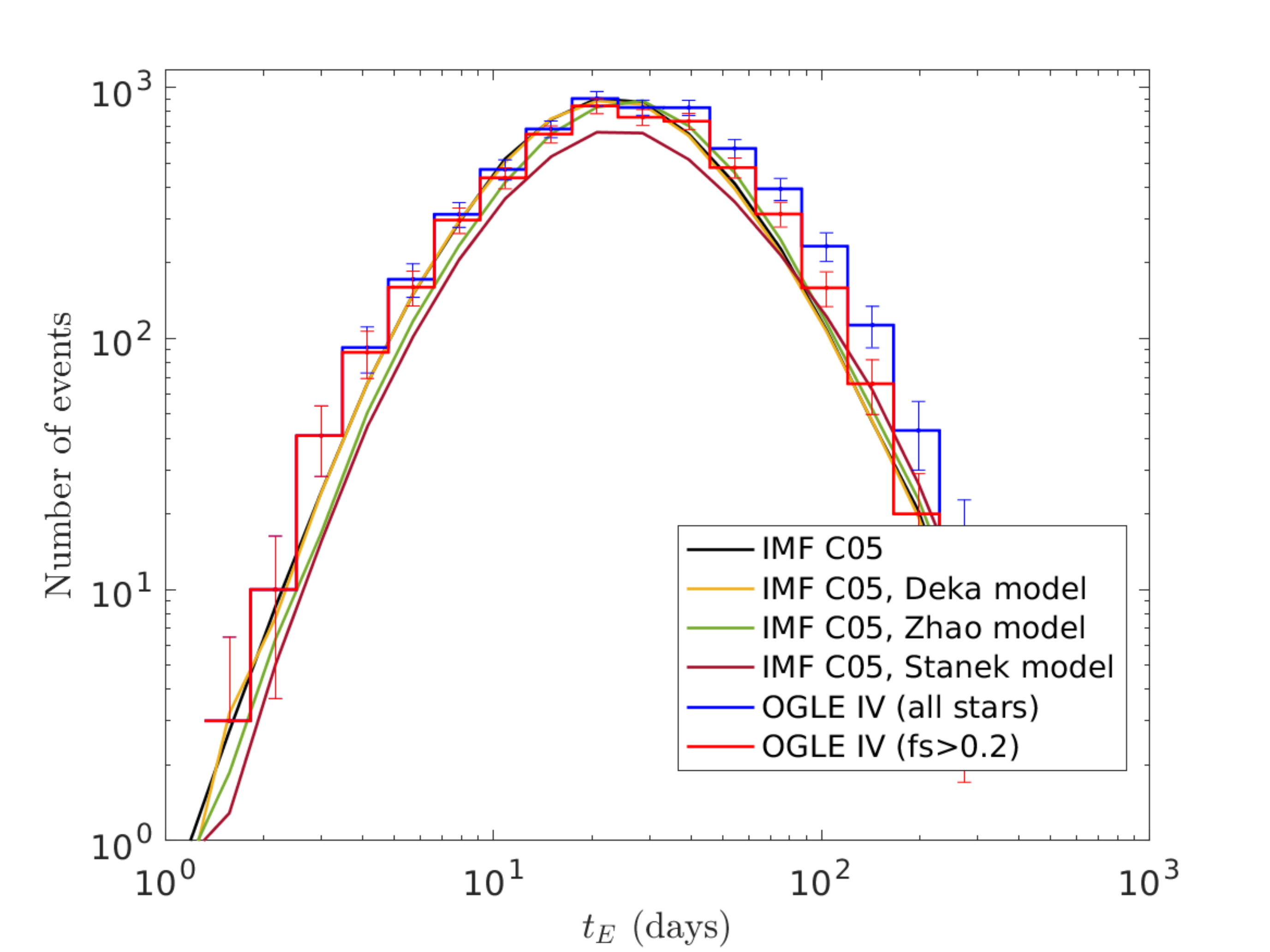}
\includegraphics[height=8.5cm,width=8.8cm]{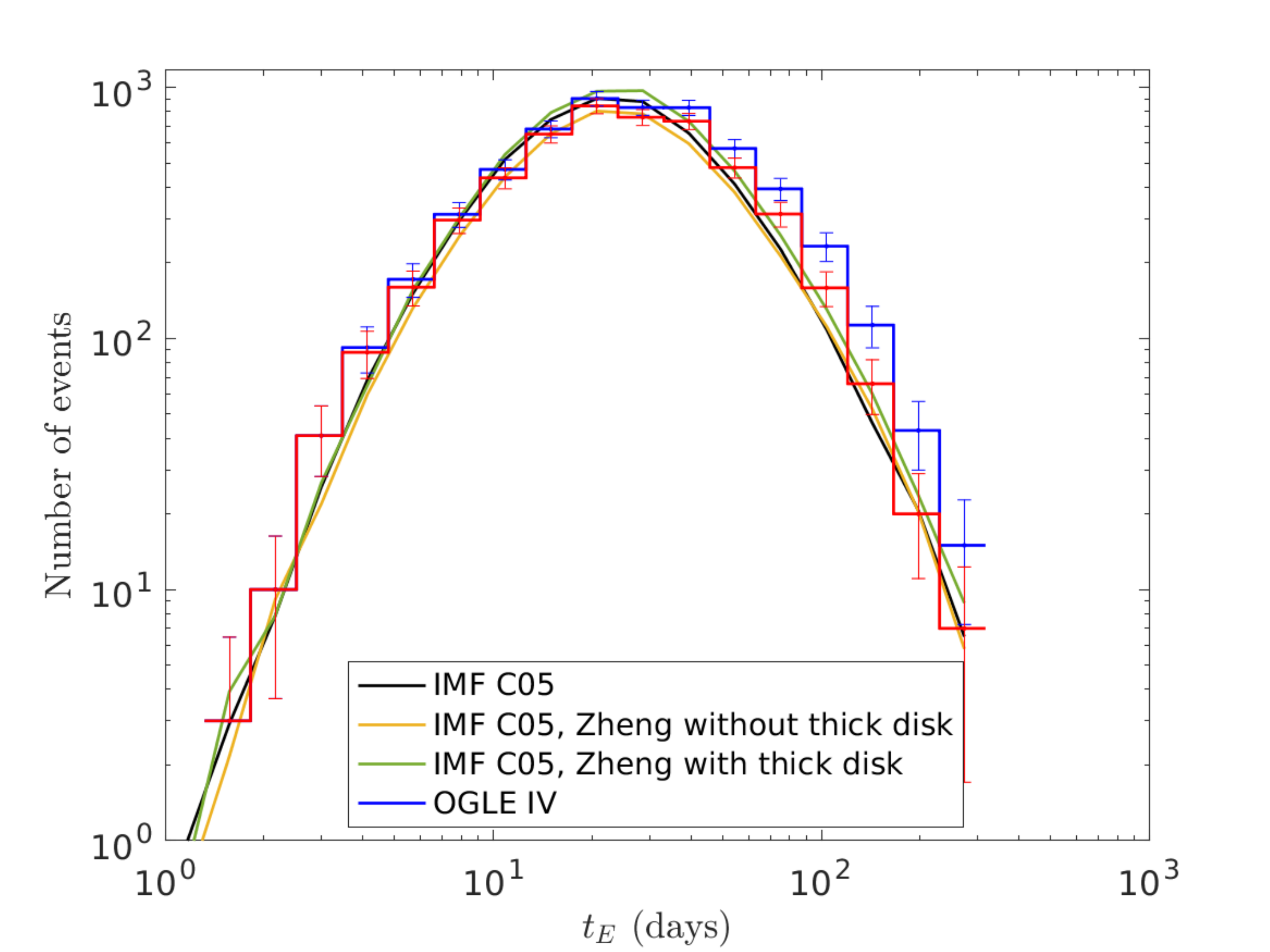}
\includegraphics[height=8.5cm,width=8.8cm]{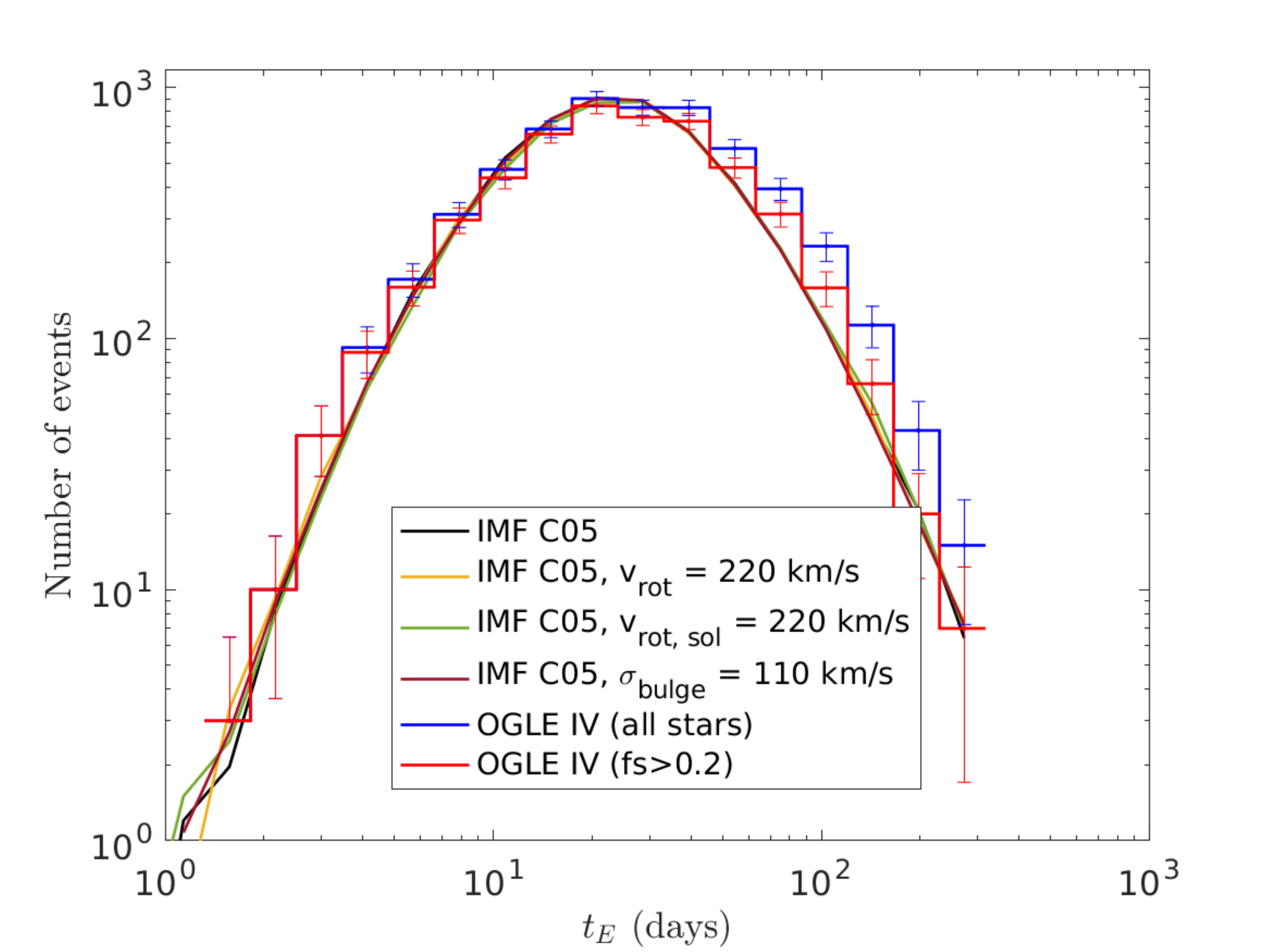}
}
\vspace{0cm}
  \caption{Influence of various parameters of the model upon the $t_{\rm E}$-histogram distribution. {\it Upper row}: left: bar angle; right: bulge model. {\it Lower row}: left: disk model; right: velocity dispersion.}
\label{fig-param}
\end{figure}

\begin{figure}[h!]
\vspace{1cm}
\center{
\includegraphics[height=5.9cm,width=5.9cm] {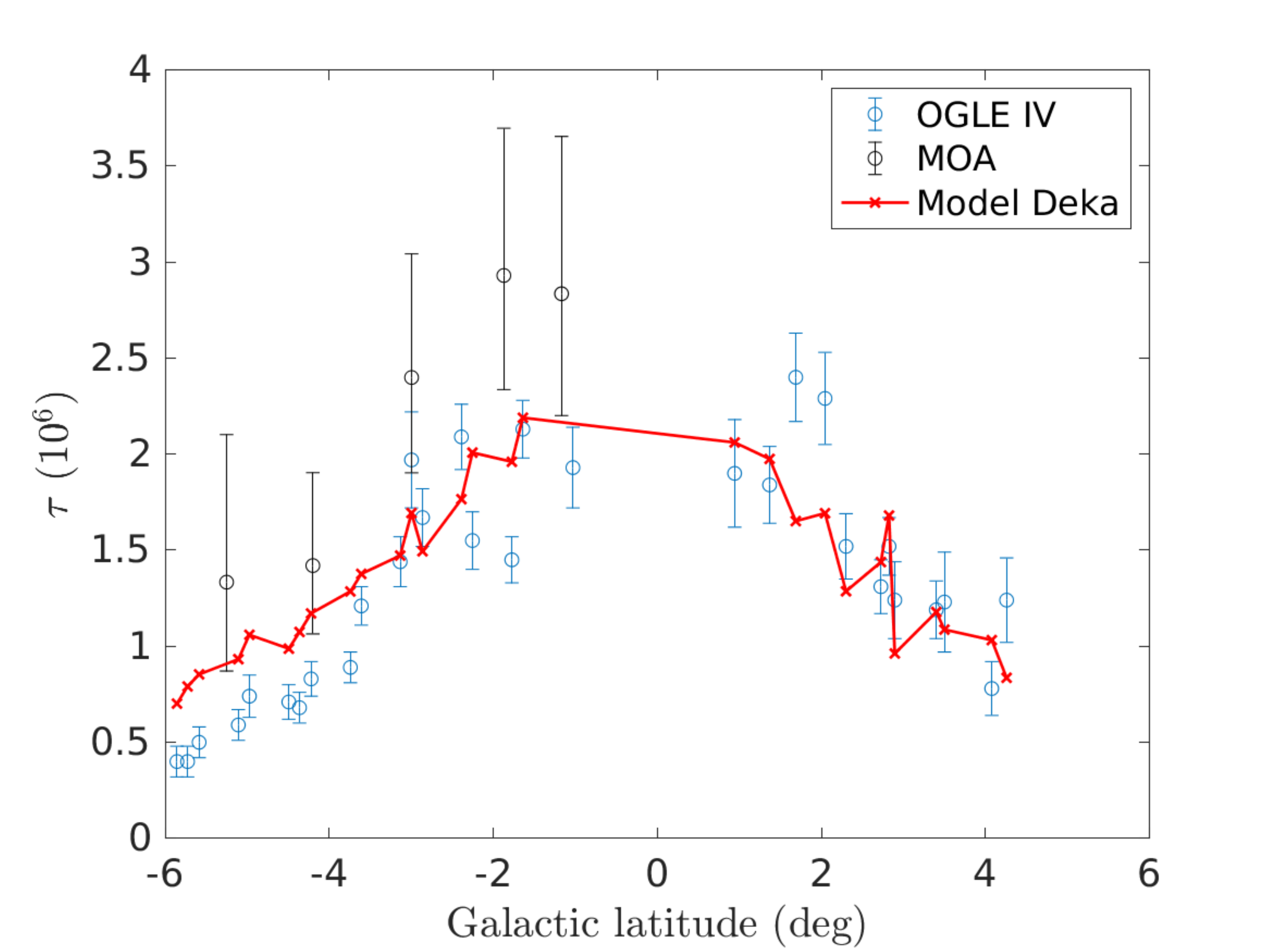} 
\includegraphics[height=5.9cm,width=5.9cm] {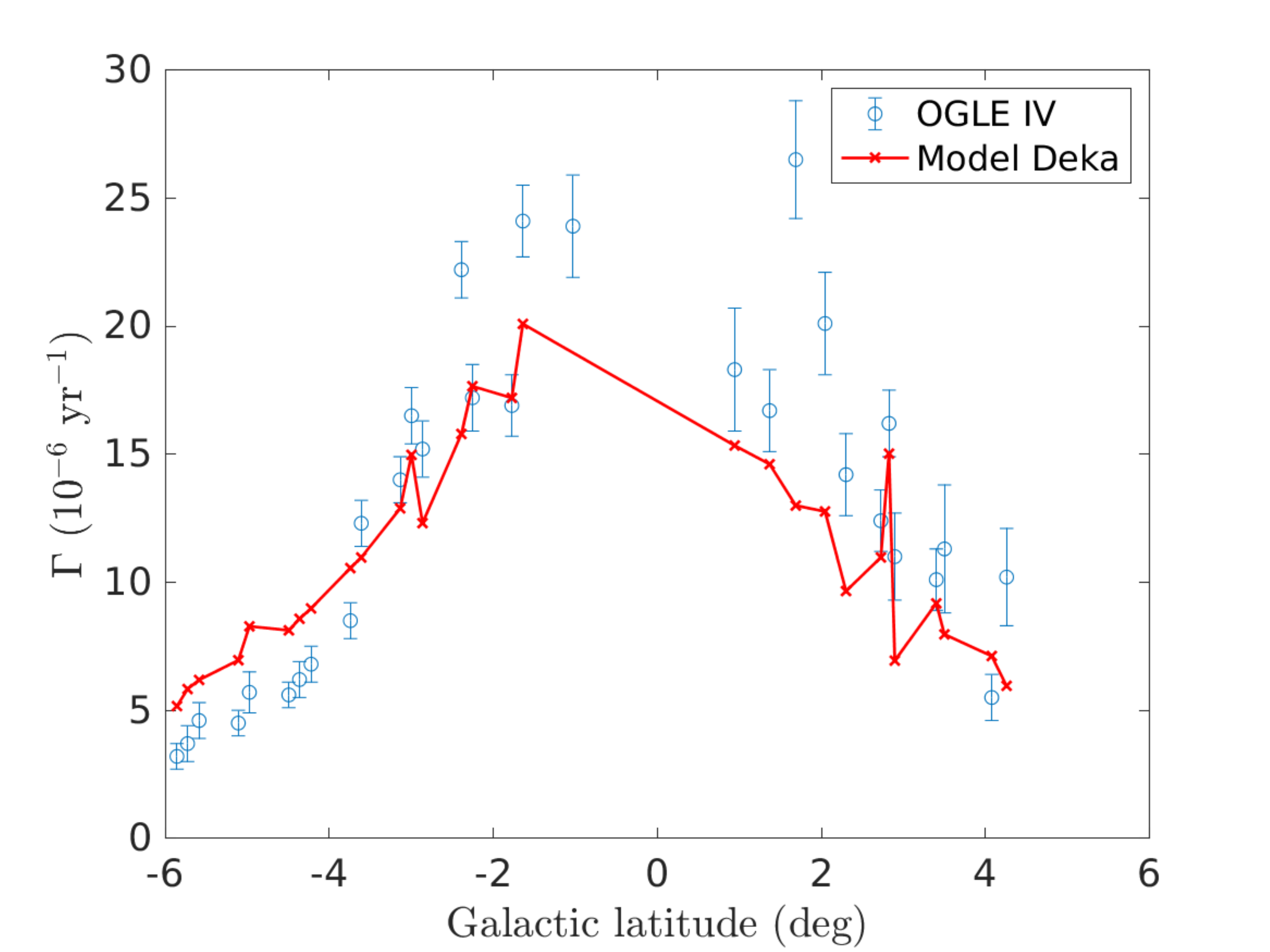} 
\includegraphics[height=5.9cm,width=5.9cm] {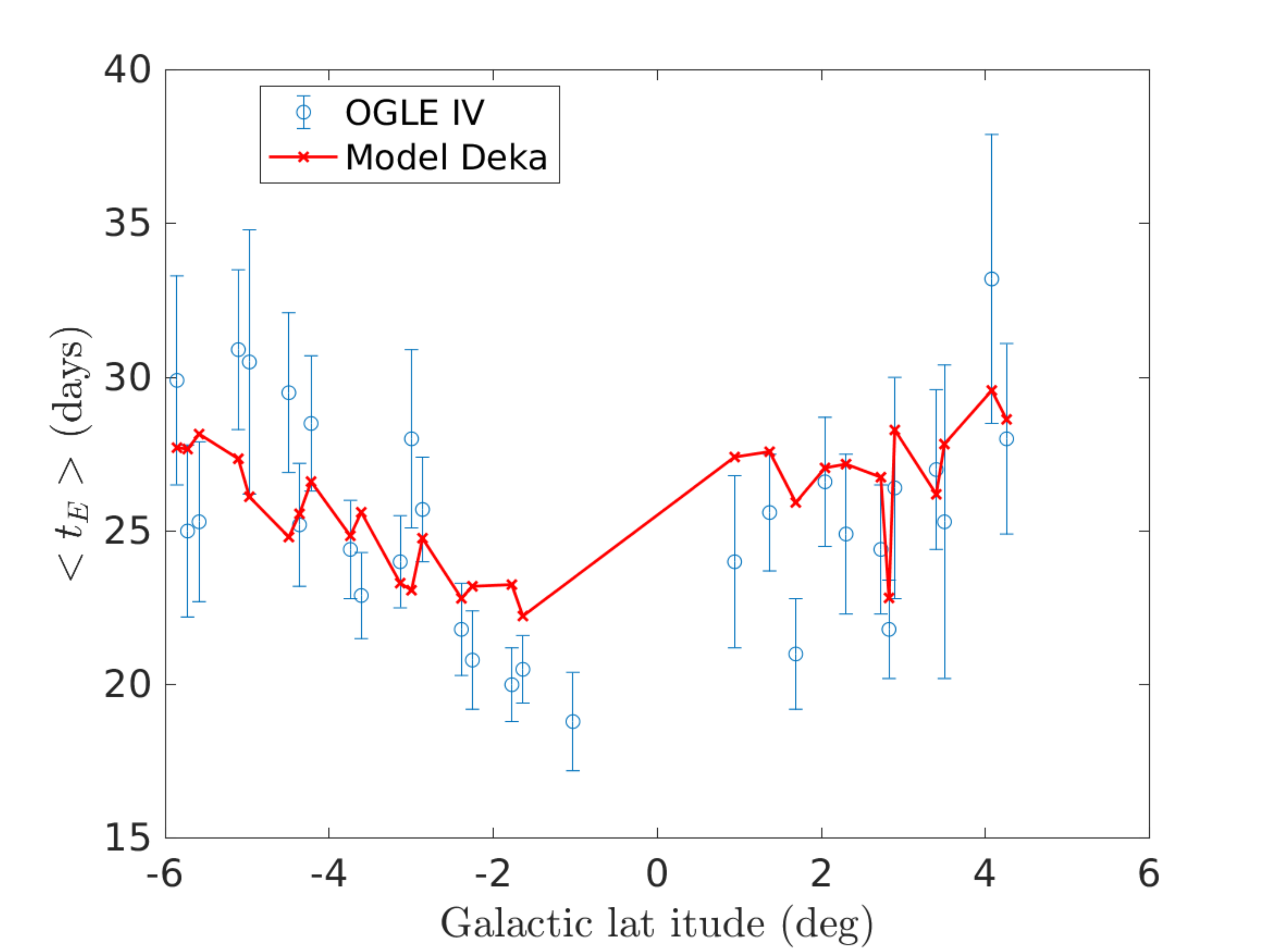} 
}
  \caption{Optical depth, $\tau$, event rate, $\Gamma$ and average characteristic time $\langle \te \rangle$, with the model based on $\delta$ Scuti stars (Deka et al. 2022), with a C05 IMF.  Calculations with $\beta=-1.0$ in Equation (\ref{nu_s}) yield similar results.}
\label{fig-Deka}
\end{figure}

\subsection{C1. Geometry}

 Figure \ref{fig-param} examines the event distribution for 2 values, $\phi=20^0$ and $35^o$, i.e., $\pm 25\%$ around our fiducial value. 
  The smaller the bar angle, the larger the mass along the l.o.s and then the larger the optical depth and the number of events. 
As seen from the values of $\tau$ and $\Gamma$ (Table \ref{para}) and the timescale distribution, the agreement of our fiducial model and of the recent model of Deka et al. (2022) (see \S\ref{Deka} below) with the data suggests a Galactic bar around our fiducial angle $\phi \simeq 28^o$.

\subsection{C2. Bulge and Disk Models}
\label{model}

The choice of the bulge and the disk models  affects the optical depth and the event rate, a direct consequence of the different total mass for each model. Figure \ref{fig-param} and Table \ref{para} compare the results obtained with our fiducial model and the ones of Stanek et al. (1997), Zhao (1996) and Deka et al. (2022) for the bulge and Zheng et al. (2001) for the disk. For the latter case, the figure also highlights the contribution of the thick disk. As seen in the figure and in Table \ref{para}, none of these models provides as good an agreement with the data as our fiducial model. However, while the Stanek model is clearly excluded,
the Zheng and Zhao ones are marginally compatible with the observations. It is reassuring to see that the uncertainties in the disk and bulge Galactic models thus seem to be modest.
 
\subsection{C3. Velocity distribution}
\label{velocity}
The optical depth does not depend on the velocity distribution and thus does not vary with the latter. In contrast, the velocity distribution affects not only the event rate but also the event distribution itself. Figure \ref{fig-param} and Table \ref{para} present the results for different disk rotation velocities and for a different velocity dispersion in the bulge. Decreasing the bulge or disk velocity dispersion decreases the number of short-time events. 
For the disk, we have examined (i) a case $\Theta_0=220\,\kms$ for the LSR normalization in Equation(\ref{vrot}) and (ii) a case $V_{rot}(R)={\rm constant}=220\,\kms$. We have also examined a case $\sigma_{bulge}=\sigma_{\rm BW}={\rm constant}=110\,\kms$ for the bulge. As seen in the figure, the impact on the event distribution remains almost negligible. 

\subsection{C4. Density of Source Stars}
\label{source}

The choice of the density of source stars, $\rho_s$, influences the distribution of $\te$. 
Some Galactic surveys  toward the bulge only include Red Clump Giants (RCG) as sources. These are very localized ($D_s=8.5\pm10\%$ kpc, Moniez (2010)). When considering such observations, we take $\rho_s(D_S)=\rho_{bulge}(D_S)$ for the source density and
$\rho_L(D_S)=\rho_{disk}(D_S)+\rho_{bulge}(D_S)$ for the lens density, and $\beta=0$ in Equation(\ref{nu_s}). For the general case, we take
$\rho_s=\rho_{disk}+\rho_{bulge}$ in both cases (see \S\ref{math} and \S\ref{source_stars}). 
The impact of the parameter $\beta$ upon $\tau$, $\Gamma$, $\langle \te \rangle$ and the timescale distribution is examined in \S\ref{tau-gamma} and  \S\ref{OGLE-all}.
A larger value of $\beta$ implies that more sources are visible, which increases the number of expected events and the optical depth. 
A value $\beta=-1.2$ yields a nearly perfect agreement with the timescale histogram  with $f_s>0.2$ and  thus has been taken as our fiducial value.
Note that the fact that $\Gamma$ is underestimated for the central fields with a C05 IMF (see Fig. \ref{fig-tau}) is not surprising, since, as examined in \S\ref{central}, the IMF in the central fields departs for this IMF.

\subsection{C5. Model based on $\delta$ Scuti stars}
\label{Deka}

Recently, Deka et al. (2022) have derived a 3D structure of the bulge using OGLE-IV $\delta$ Scuti stars. The bar in this model is more inclined than that for our fiducial model, with $\theta=22^o$
instead of 28$^o$, and is more compact, with normalized ($a\equiv 1$) axes ratios $(a:b:c)=1,0.348,0.421$. The $\tau$, $\Gamma$, $\langle \te \rangle$ values
and the event
timescale histogram distribution  obtained with this model, with the C05 IMF, are displayed in Table \ref{para} and Figures \ref{fig-param} and \ref{fig-Deka}. As seen from this figure and Figure  \ref{fig-tau}, the agreement with the OGLE-IV data for $\tau$ and $\Gamma$ is not as good as the one obtained with our fiducial model: we note the too-large values of $\tau$ and $\Gamma$ obtained with this model, notably in the peripheral fields, a consequence of the more inclined and more compact bar.  
\bigskip

To conclude this section, it seems fair to say that the global uncertainty in our Galactic model can be considered as rather modest. A more thorough (combined) exploration of the different model parameters around our fiducial values, involving a 3D model, would probably yield a perfect agreement with all the data. However, this goes beyond the present study, whose aim is to explore the impact of the IMF.

\begin{figure}[h!]
\vspace{1cm}
\center{
\includegraphics[height=8cm,width=8cm] {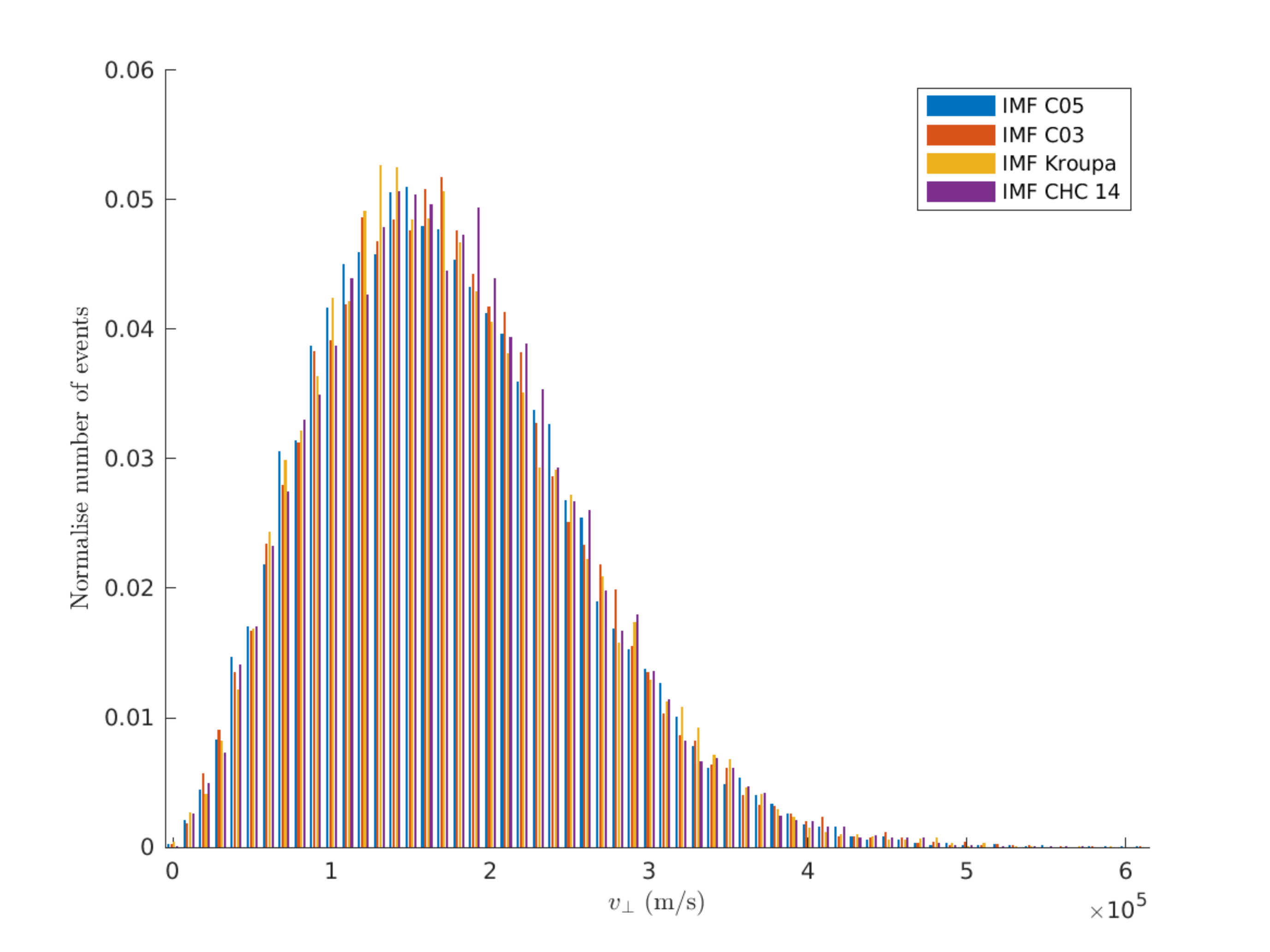} 
\includegraphics[height=8cm,width=8cm] {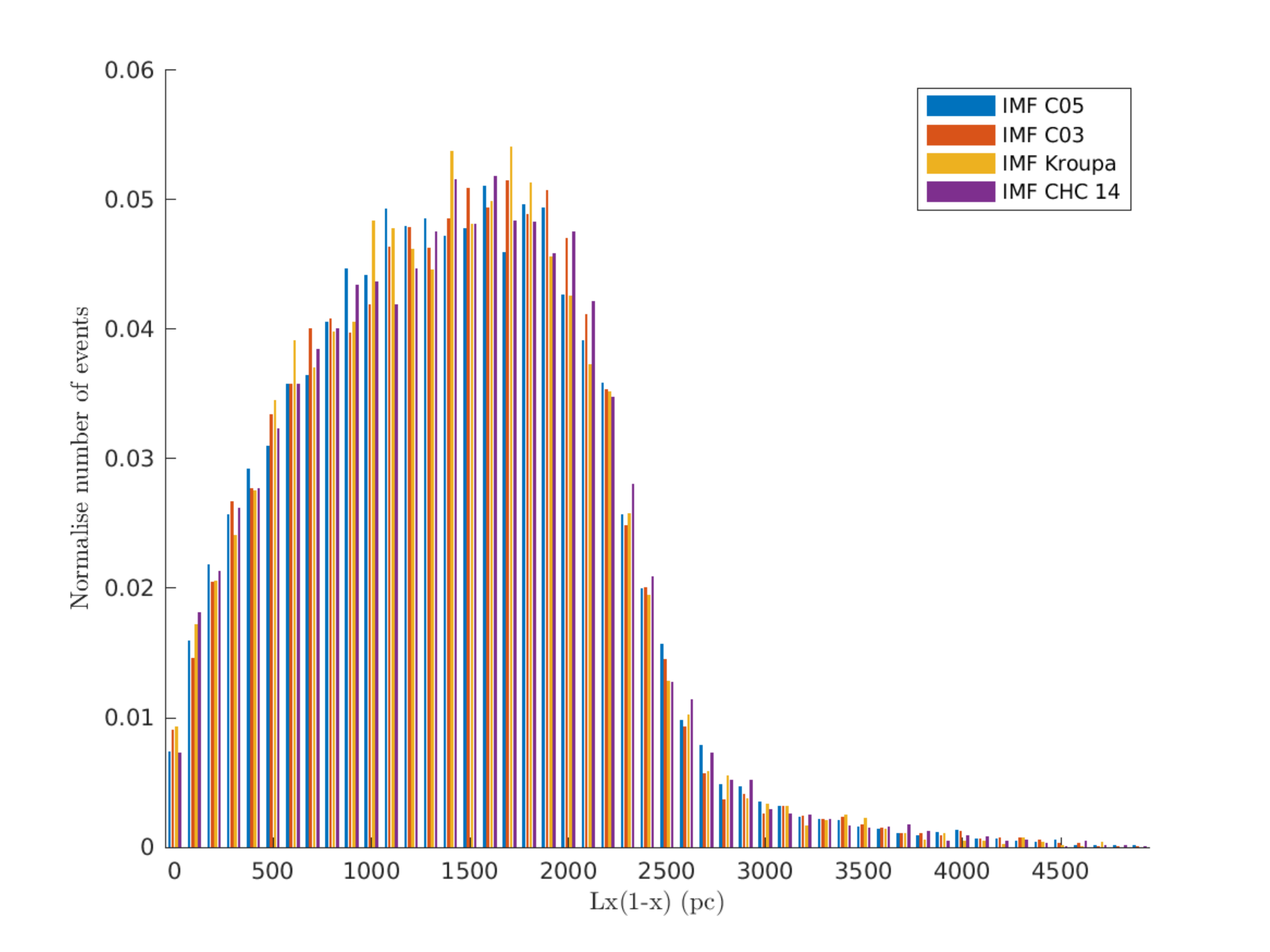} 
}
  \caption{Comparison of  the normalized number of events, without taking into account the experimental efficiency, as a function of
$v_\perp$ and $D_Sx(1-x)$ for the 4 IMFs examined in \S\ref{central}. }
\label{fig-vperp}
\end{figure}

\section{D. The distributions of transverse velocity and lens-source distance modulus in the central fields}
\label{central-fields}

Figure \ref{fig-vperp} portrays the statistical distributions of the number of events for the OGLE-IV central fields survey as a function of $v_\perp$ and $D_Sx(1-x)$, respectively, before taking into account the experimental efficiency, for the 4 IMFs examined in \S\ref{central}. As seen in the figure, the distributions are quite similar for all IMFs, demonstrating the fact that the effective probabilities $P_{{eff}}(x)$ and $P_{{eff}}(v_\perp)$ (eqns.(\ref{Px})(\ref{Pv})) do not depend on the mass. This ensures the validity of our Monte Carlo calculations. The atypical  distribution in Figure \ref{fig-central} for the central fields
thus really stems from a different underlying IMF.

\end{document}